\documentclass[journal=aaembp,manuscript=article]{achemso}

\usepackage{chemformula} 
\usepackage[T1]{fontenc} 
\usepackage{pdfpages}

\author{Zimeng Guo}
\affiliation[Kyushu University Chikushi]
{Department of Applied Science for Electronics and Materials, Kyushu University, Fukuoka 816-8580, Japan}
\alsoaffiliation[CREST Japan]
{Core Research for Evolutionary Science and Technology, Japan Science and Technology Agency, Tokyo 102-0076, Japan}

\author{Hongye Gao}
\affiliation[Kyushu University URC]
{The Ultramicroscopy Research Center, Kyushu University, Fukuoka 819-0395, Japan}

\author{Keisuke Kondo}
\affiliation[Nagoya University]
{Department of Materials Physics, Nagoya University, Nagoya 464-8603, Japan}

\author{Takafumi Hatano}
\affiliation[Nagoya University]
{Department of Materials Physics, Nagoya University, Nagoya 464-8603, Japan}
\alsoaffiliation[CREST Japan]
{Core Research for Evolutionary Science and Technology, Japan Science and Technology Agency, Tokyo 102-0076, Japan}

\author{Kazumasa Iida}
\affiliation[Nagoya University]
{Department of Materials Physics, Nagoya University, Nagoya 464-8603, Japan}
\alsoaffiliation[CREST Japan]
{Core Research for Evolutionary Science and Technology, Japan Science and Technology Agency, Tokyo 102-0076, Japan}

\author{Jens H\"anisch}
\affiliation[Karlsruhe Institute of Technology]
{Institute for Technical Physics, Karlsruhe Institute of Technology, Hermann-von-Helmholtz-Platz 1, 76344 Eggenstein-Leopoldshafen, Germany}

\author{Hiroshi Ikuta}
\affiliation[Nagoya University]
{Department of Materials Physics, Nagoya University, Nagoya 464-8603, Japan}

\author{Satoshi Hata}
\email{hata.satoshi.207@m.kyushu-u.ac.jp}
\affiliation[Kyushu University Chikushi]
{Department of Applied Science for Electronics and Materials, Kyushu University, Fukuoka 816-8580, Japan}
\alsoaffiliation[CREST Japan]
{Core Research for Evolutionary Science and Technology, Japan Science and Technology Agency, Tokyo 102-0076, Japan}
\alsoaffiliation[Kyushu University URC]
{The Ultramicroscopy Research Center, Kyushu University, Fukuoka 819-0395, Japan}
\alsoaffiliation{Department of Advanced Materials Science and Engineering, Faculty of Engineering Sciences, Kyushu University, Fukuoka 816-8580, Japan}

\title[ACS Applied Electronic Materials]
  {Nanoscale Texture and Microstructure in NdFeAs(O,F)/IBAD-MgO Superconducting Thin Film with Superior Critical Current Properties}


\begin{document}




\begin{abstract}
  This paper reports the nanoscale texture and microstructure of a high-performance NdFeAs(O,F) superconducting thin film grown by molecular beam epitaxy on a textured MgO/\ch{Y2O3}/Hastelloy substrate. The NdFeAs(O,F) film forms a highly textured columnar grain structure by epitaxial growth on the MgO template. Although the film contains stacking faults along the $ ab $-plane as well as grain boundaries perpendicular to the $ ab $-plane, good superconducting properties are measured: a critical temperature, $ T _{\rm c} $, of 46\,K and a self-field critical current density, $ J_{\rm c} $, of $ 2 \times 10^6 \,{\rm A/cm^2} $  at 4.2\,K. Automated crystal orientation mapping by scanning precession electron diffraction in transmission electron microscopy is employed to analyze the misorientation angles between adjacent grains in a large ensemble (247 grains). 99\% of the grain boundaries show in-plane misorientation angles ($ \Delta \gamma $) less than the critical angle $ \theta_{\rm c} $, which satisfies one of the necessary conditions for the high $ J_{\rm c} $. Comparing the columnar grain size distribution with the mean distance of the flux line lattice, the triple junctions of low-angle grain boundaries are found to be effective pinning centers, even at high temperatures ($ \ge 35 $\,K) and/or low magnetic fields.

\end{abstract}

\section{Introduction}
Since the discovery of Fe-based superconductors (FBS),\cite{Kamihara2008} fundamental properties and possibilities for applications of them have been investigated intensively. Among FBSs, $ Ln $FeAs(O,F) ($ Ln = lanthanide $) show the highest superconducting transition temperature, $ T_{\rm c} $ (up to around 58\,K)\cite{Fujioka2013}, which is advantageous for applications because of a large margin between $ T_{\rm c} $ and the operating temperatures, e.g. those a cryocooler can reach. Additionally, the theoretical limit of the critical current density $ J_{\rm c} $ (i.e., depairing current density $ J_{\rm d} $ for which Cooper pairs are separated by kinetic energy) is around 170\,MA/cm$ ^2 $ at zero kelvin,\cite{Pallecchi2015} which is the highest value among FBSs. Enhancing the $ J_{\rm c} $ of polycrystalline $ Ln $FeAs(O,F) makes this material more attractive for applications.\cite{Haenisch2019} However, polycrystalline $ Ln $FeAs(O,F) generally does not show the best performance, because $ J_{\rm c} $ significantly decreases in the presence of high-angle grain boundaries (HAGB) with misorientation angles larger than the critical angle $ \theta_{\rm c} $, for which inter-grain $ J_{\rm c} $ starts to decrease exponentially. Hence, these HAGBs become serious weak links.\cite{Durrell2011,Iida2020} In our previous study,\cite{Iida2019} $ \theta_{\rm c} $ was evaluated to be $ 8.5^{\rm \circ} $ for NdFeAs(O,F) bicrystals with symmetrical [001]-tilt boundaries, which is close to $ \theta_{\rm c} \sim 9^{\rm \circ} $ for Co-doped \ch{BaFe2As2} (Ba-122)\cite{Katase2011} and Fe(Se,Te) bicrystals.\cite{Si_2015,Sarnelli2017} Therefore, strict control of the texture is required to fully utilize its potential advantages in applications. \par

Furthermore, it is promising that the grain boundary engineering can enhance in-field $ J_{\rm c} $ for some superconductors. For example, a P-doped Ba-122 film on a poorly textured polycrystalline MgO template exhibited notably better magnetic flux pinning properties and, therefore, higher $ J_{\rm c} $ values than such a film on a well-textured MgO template.\cite{Sato2016} This result was explained by a higher density of low-angle grain boundaries (LAGB) in the poorly textured film, which act as pinning centers. This kind of grain boundary pinning was also reported for other superconductors, such as \ch{MgB2}.\cite{Kitaguchi2004} One can therefore presumably take advantage of introducing grain boundaries also in $ Ln $FeAs(O,F) films to pin magnetic flux lines and enhance in-field $ J_{\rm c} $ if grain boundaries are appropriately engineered, i.e. with grain boundary angles predominantly smaller than $ \theta_{\rm c} $. \par

In regard to solving the aforementioned challenge, the film growth of textured NdFeAs(O,F) by molecular beam epitaxy (MBE) on highly textured MgO templates\cite{Iida2014} has been studied. These templates are prepared by ion-beam-assisted deposition (IBAD) on \ch{Y2O3}-buffered Hastelloy tapes.\cite{Sheehan2011} IBAD is a typical technique for depositing a biaxially textured film on an untextured metal tape.\cite{Iijima1992} After optimization of various fabrication parameters, the NdFeAs(O,F) thin film exhibited a self-field $ J_{\rm c} $ of $ 10^6 \,{\rm A/cm^2} $ at 4.2\,K, which is comparable to $ J_{\rm c} $ for single-crystalline NdFeAs(O,F) thin films. Because NdFeAs(O,F) thin films on textured MgO templates are expected to contain a high density of grain boundaries and other defects that influence $ J_{\rm c} $, it is worthwhile to investigate the microstructure of the NdFeAs(O,F)/IBAD-MgO thin film in fine detail. \par

In this paper, the microstructure of this NdFeAs(O,F)/IBAD-MgO thin film is analyzed by electron microscopy. Atomic-scale microstructural features and the nanoscale crystal orientation distribution have been evaluated by several TEM and scanning TEM (STEM) techniques. Especially, automated crystal orientation mapping (ACOM) using scanning precession electron diffraction (SPED, see also Supporting Information (SI)) has revealed quantitative and quasi-statistical information about the nanoscale texture and microstructure. They are correlated with the macroscopic transport properties: the misorientation angles between neighboring grains compared with $ \theta_{\rm c} $ as well as the grain size distribution compared with the mean flux line distance. Based on the assessments described above, the essential features of the microstructure are directly correlated to the superior critical current properties of the NdFeAs(O,F)/IBAD-MgO thin film. \par

\section{Results and discussion}

\subsection{Crystal structure and global texture}

\begin{figure}[htb]
  \includegraphics[width=\textwidth]{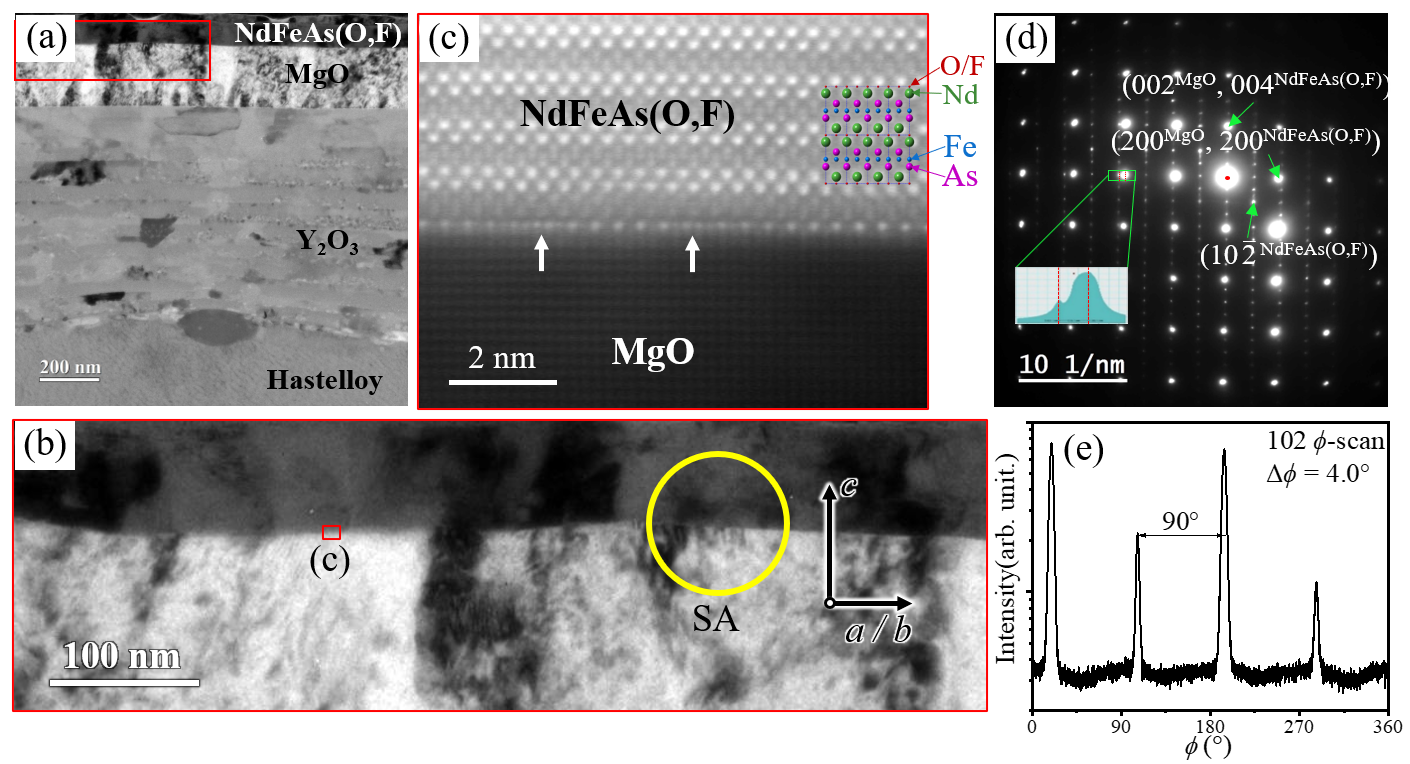}
  \caption{(a) Cross-sectional bright-field TEM image of the highly textured NdFeAs(O,F) thin film grown by MBE on IBAD-MgO/\ch{Y2O3}/Hastelloy. (b) Enlarged image of the red rectangular region in (a). (c) Atomic-resolution ADF-STEM image of the interface area between NdFeAs(O,F) and MgO. (d) Selected-area electron diffraction (SAED) at the yellow circular area (SA) in (b). (e) The $ hkl = 102 $  $ \phi $-scan of NdFeAs(O,F) shows four-fold symmetry and no $ 45^{\rm \circ} $ in-plane rotated texture components were observed.}
  \label{fgr:1}
\end{figure}

The bright-field TEM image shown in Figure \ref{fgr:1}a is a cross-sectional view of the film sample. A columnar contrast is observed in both NdFeAs(O,F) and MgO (Figure \ref{fgr:1}b). The azimuthal $ \phi $-scan of the NdFeAs(O,F) $ hkl = 102 $  peak obtained by X-ray diffraction (XRD) (Figure \ref{fgr:1}e) indicates a strong biaxial texture. The average full width at half maximum (FWHM) of the peaks, $ \Delta \phi = 4.0^{\rm \circ} $, is well below $ \theta_{\rm c} $. The epitaxial relationship as confirmed by electron diffraction (Figure \ref{fgr:1}d) is $ (001)[100]_{\rm NdFeAs(O,F)} \parallel (001)[100]_{\rm MgO} $. Additionally, the atomic-resolution annular dark-field (ADF) STEM image (Figure \ref{fgr:1}c) clearly exhibits the layered atomic structure of the NdFeAs(O,F) film: the FeAs conductive layers and the Nd(O,F) charge-reservoir layers alternate along the crystallographic $ c $ direction. The white arrows point out misfit dislocations at the interface, which relieve the inner stress induced by lattice mismatch between NdFeAsO ($ a = 0.399 $\,nm)\cite{Kito2008} and MgO ($ a = 0.421 $\,nm)\cite{SASAKI1979}.

\subsection{Grain size distribution}

\begin{figure}[htbp]
	\includegraphics[width=\textwidth]{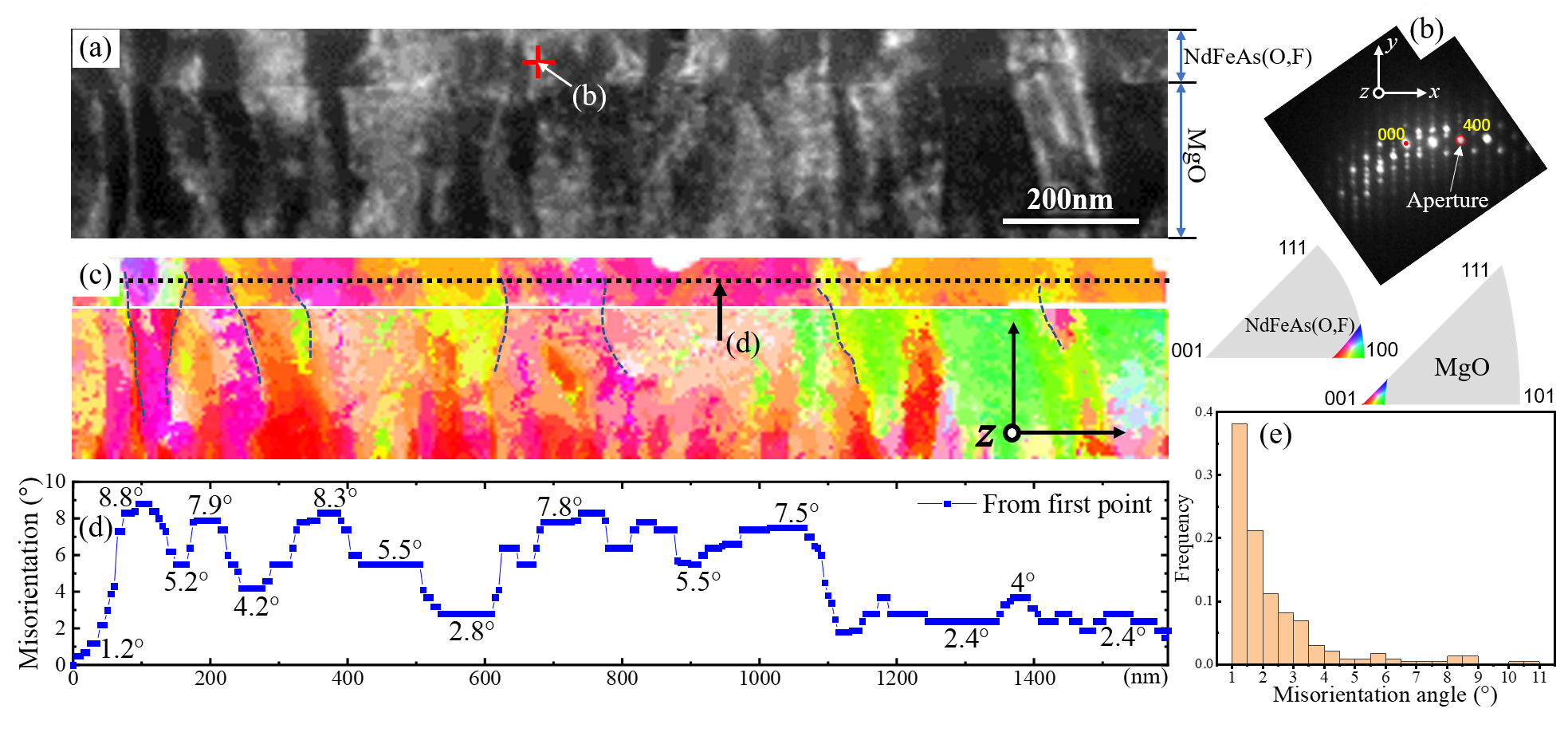}
	\caption{Scanning precession electron diffraction (SPED) data measured in an area of 300\,nm height and 1600\,nm width. (a) Virtual dark-field (DF) TEM image of the NdFeAs(O,F) thin film and the MgO template exported from the SPED data. The diffraction intensities at $ hkl = 400 $ marked in (b) were used for displaying the virtual DF-TEM image in (a). (b) Precession electron diffraction pattern acquired at the red cross location in (a). (c) Nanoscale crystal orientation map on $ z $-axis view direction, and the corresponding color codes. (d) The typical misorientation angle profile along the dotted horizontal line in (c) with respect to the first point. (e) The distribution of misorientation angles between adjacent grains in 247 grains acquired from 10 random TEM foils. Only the misorientation angles larger than $ 1^{\rm \circ} $ were counted.}
	\label{fgr:2}
\end{figure}

\begin{figure}[htbp]
	\includegraphics[width=0.5\textwidth]{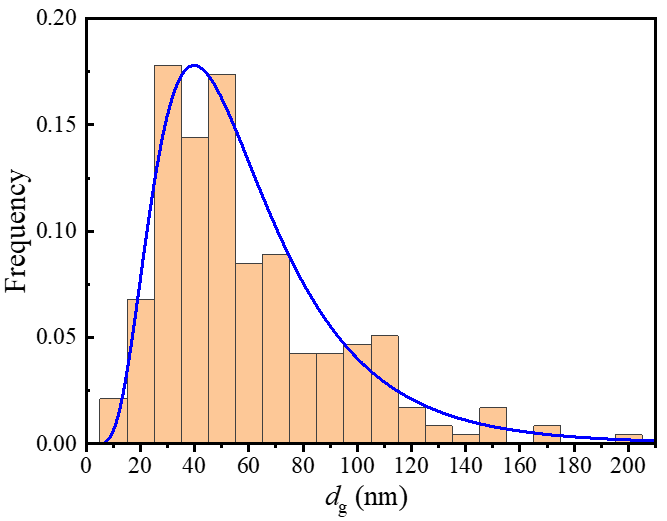}
	\caption{The statistical histogram of the distances between adjacent grain boundaries, $ d_{\rm g} $, and the relevant log-normal distribution curve. The mean value of $ d_{\rm g} $ was calculated as 61\,nm.}
	\label{fgr:3}
\end{figure}

Figure \ref{fgr:2}a shows a virtual dark-field TEM image of the NdFeAs(O,F) film and the MgO template exported from an SPED data set using the intensity distribution at the $ \textit{\textbf{g}} (hkl) = 400 $ diffraction spot (Figure \ref{fgr:2}b). A nearly vertical stripe contrast is recognized in both \mbox{NdFeAs(O,F)} and IBAD-MgO, which is continuously connected between the two phases. This image contrast indicates the texture transfers from the MgO template to the NdFeAs(O,F) film, and thus, the texture of the film can be engineered by this kind of technical substrates. The nanoscale crystal orientation map in Figure \ref{fgr:2}c (see also Figure S2 in SI) shows the distribution of strongly $ c $-axis correlated grain boundaries and the texture in great detail, revealing a wide distribution of grain sizes (about 20--120\,nm). Figure \ref{fgr:2}d shows a typical misorientation angle profile along the dotted line in the NdFeAs(O,F) film in Figure \ref{fgr:2}c, which was calculated between each data point and the left-edge point on the line. This method enables us to measure precisely the misorientation angle between adjacent grains, $ \Delta g $, and the distance between adjacent grain boundaries, $ d_{\rm g} $. Figure \ref{fgr:2}e shows the distribution of $ \Delta g $ calculated from 247 grains acquired from ten random cross-sectional foil samples. These $ \Delta g $ values are very small and show a highly compact distribution (peaked at about $ 1.2^{\rm \circ} $), which means that the majority of the grain boundaries in NdFeAs(O,F) are LAGBs or sub-GBs. In addition, the distribution of $ d_{\rm g} $ measured from 247 grains, Figure \ref{fgr:3}, shows a mean value of 61\,nm, and grains in the $ d_{\rm g} $ range of 30--55\,nm account for a large proportion. This will be discussed together with the matching-field effect.

\subsection{Superconducting properties}

\begin{figure}[htb]
	\includegraphics[width=0.7\textwidth]{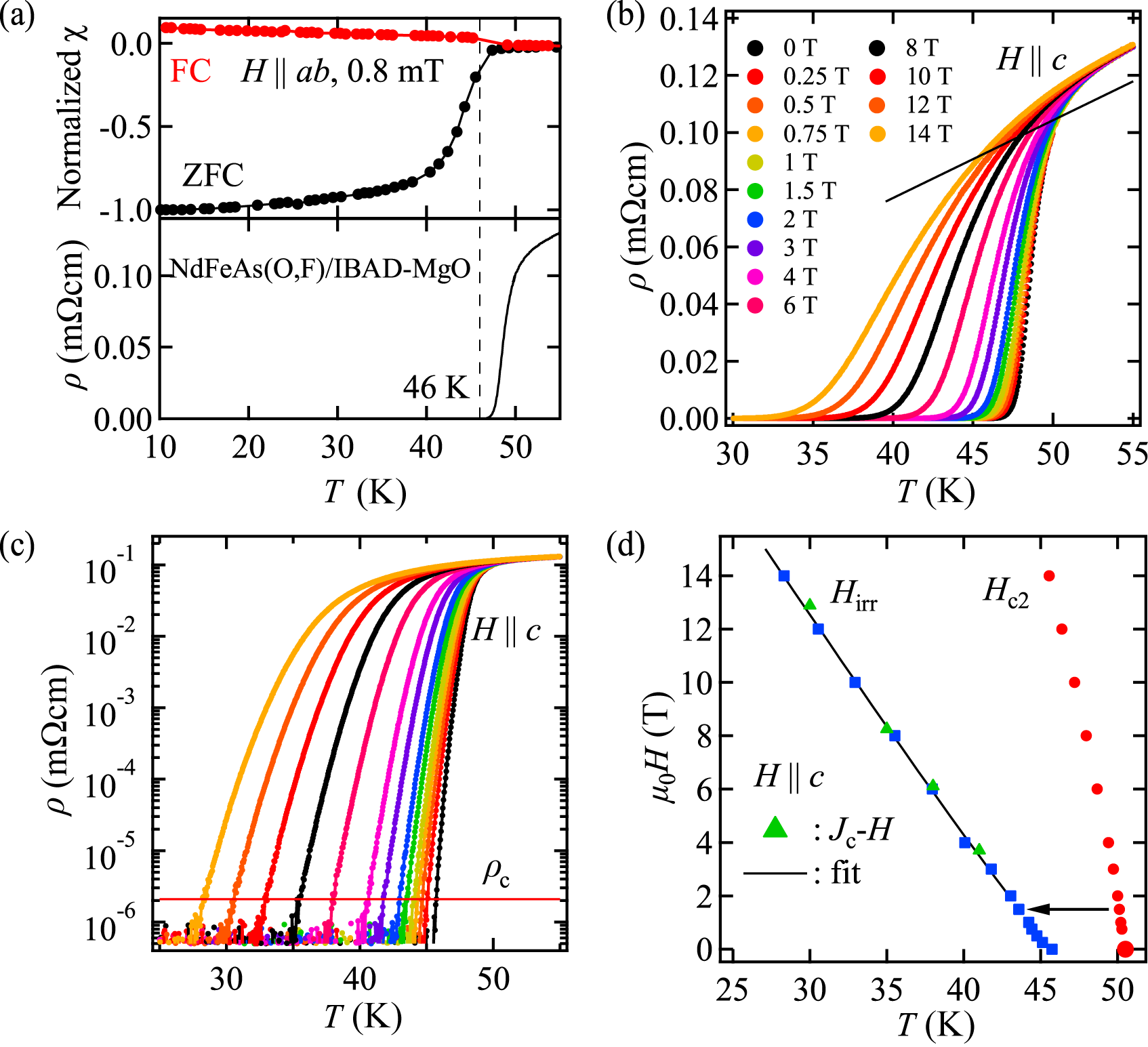}
	\caption{(a) Temperature ($ T $) dependence of the magnetization for zero field-cooled (ZFC) and field-cooled (FC) branches, and the resistivity ($ \rho $) of the NdFeAs(O,F)/IBAD-MgO sample. A small magnetic field of 0.8\,mT was parallel to the $ ab $-plane for magnetization measurements. The magnetization curves were normalized to the value at 10\,K. The onset transition temperature ($ \sim 46$\,K) of the magnetization agrees with the zero-resistivity temperature. (b) In-field $ \rho $-$ T $ measurements for $ H \parallel c $. (c) Semi-logarithmic plots of (b) for determining $ H_{\rm irr} $, where $ \rho_{\rm c} $ is the $ H_{\rm irr} $ criterion. (d) Magnetic phase diagram of NdFeAs(O,F)/IBAD-MgO for $ H \parallel c $. The data determined from $ J_{\rm c} $-$ H $ measurements (as shown by triangles) follow well the $ H_{\rm irr} $ line. The arrow indicates the matching field. The black fitting line shows a power law relation $ \mu_0 H_{\rm irr} \sim (1 - T / T_{\rm c})^n $ with $ T_{\rm c} = 46 $\,K and $ n = 1.1 $.}
	\label{fgr:4}
\end{figure}

\begin{figure}[htb]
	\includegraphics[width=0.7\textwidth]{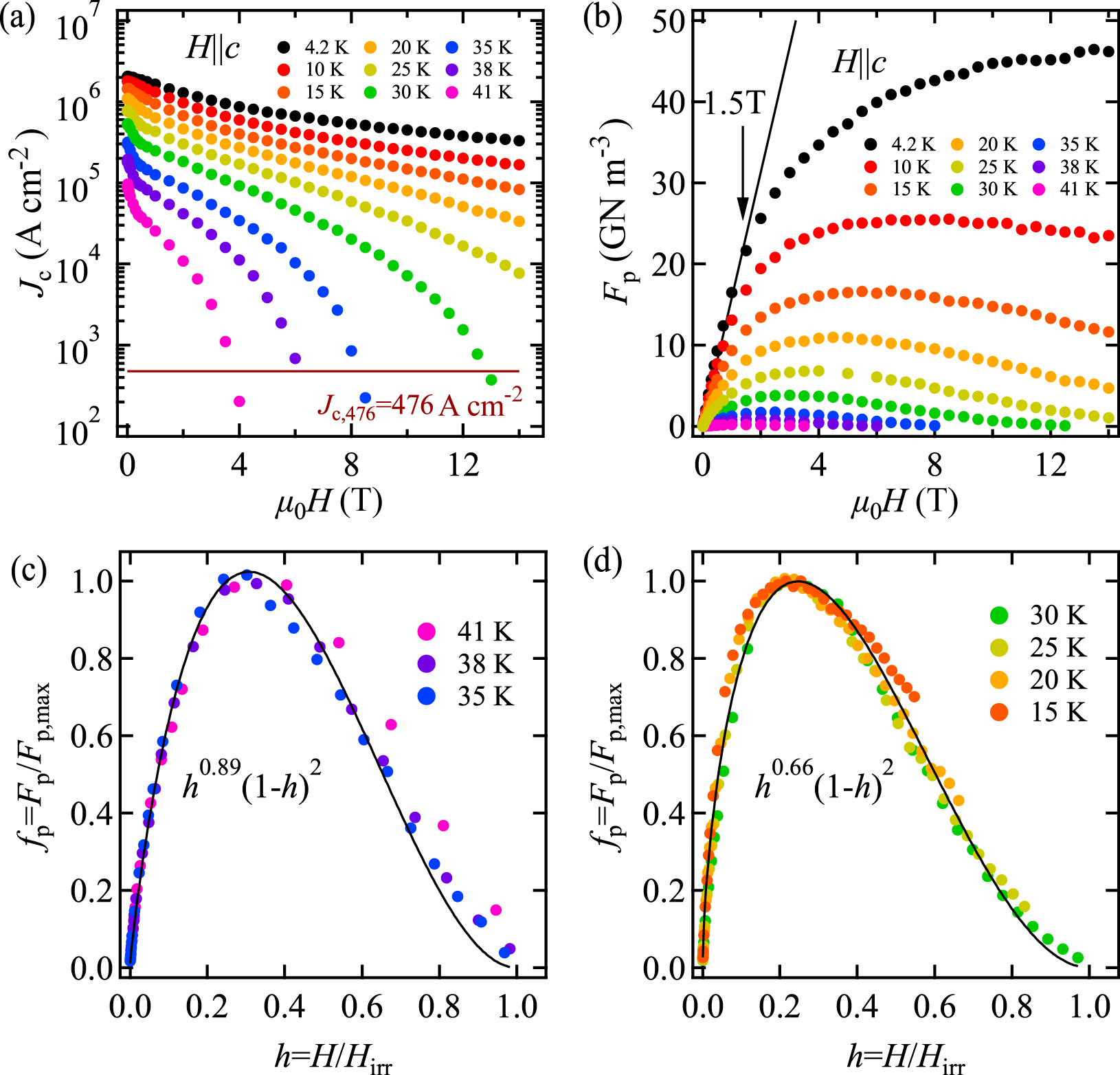}
	\caption{(a) Magnetic field ($ H $) dependence of critical current density $ J_{\rm c} $, and (b) corresponding pinning force density $ F_{\rm p} $ of the textured NdFeAs(O,F) film for $ H \parallel c $ at various temperature up to 41\,K. A criterion of 476\,A/cm$ ^2 $ for determining $ H_{\rm irr} $ is also superimposed in (a). (c), (d) Normalized pinning force density $ f_{\rm p} $. The black solid lines are master curves of $ h^p (1 - h)^2 $.}
	\label{fgr:5}
\end{figure}

The superconducting transition temperature, $ T_{\rm c} $, of this highly textured NdFeAs(O,F) film was determined as around 46\,K from the magnetization measurement, corresponding to the zero resistivity temperature, as shown in Figure \ref{fgr:4}a. Upper critical field $ H_{\rm c2} $ and irreversibility field $ H_{\rm irr} $ were determined from the temperature dependencies of the resistivity $ \rho $-$ T $ at constant applied fields $ H \parallel c $ (Figure \ref{fgr:4}b,c) and compiled to the corresponding phase diagram (Figure \ref{fgr:4}d). $ H_{\rm irr} $ values evaluated from $ J_{\rm c} $-$ H $ measurements (Figure \ref{fgr:5}a) almost overlap with those from the resistivity. Furthermore, the slope of $ H_{\rm irr} $ changes at $ \sim 1.5 $\,T (marked by arrow), which indicates the matching-field effect that will be discussed later. Above the matching field, $ H_{\rm irr} $ shows a near-linear power law relation
\begin{equation}
  \mu_0 H_{\rm irr} \sim (1 - \frac{T}{T_{\rm c}})^n 
  \label{eqn:1}
\end{equation} 
with $ T_{\rm c} = 46 $\,K and $ n = 1.1 $ (see also Figure S3a). \par

The $ J_{\rm c} $ of this NdFeAs(O,F)/IBAD-MgO film exceeds the required level for applications $ 10^5 \,{\rm A/cm^2} $ even at 14\,T and 10\,K (Figure \ref{fgr:5}a). The self-field $ J_{\rm c} $ reaches $ 2 \times 10^6 \,{\rm A/cm^2} $ at 4.2\,K, which is remarkably higher than previously reported ($ 7 \times 10^4 {\rm A/cm^2} $ at 5\,K)\cite{Iida2014} despite the similar value of $ \Delta \phi \sim 3.4^{\rm \circ} $ for the main texture component of NdFeAs(O,F). This result suggests that the presence of even a small amount of $ 45^{\rm \circ} $ in-plane rotated grains limited $ J_{\rm c} $ in the previous sample due to the erosion of grain boundaries by excess of fluorine. Note that the presence of such rotated grains in Co-doped Ba-122 does not reduce $ J_{\rm c} $ because of the absence of fluorine.\cite{Haenisch2014} Figure \ref{fgr:5}b shows the corresponding pinning force densities $ F_{\rm p} $ as a function of the magnetic field, and Figure \ref{fgr:5}c,d the same data normalized as $ f_{\rm p} = F_{\rm p} / F_{\rm p,max} $ vs. $ h = H / H_{\rm irr} $, where $ H_{\rm irr} $ was estimated from Equation \ref{eqn:1} for the lowest temperatures ($ T \le 25 $\,K). For $ T \ge 35 $\,K, $ f_{\rm p} $ can be scaled with an exponent of $ p = 0.89 $ for $ h^p (1 - h)^2 $, which is close to 1. This indicates that point-like core pinning plays an important role for $ J_{\rm c} $,\cite{Iida2017,Tarantini2016} in the sense of single-occupied defects larger than coherence length $ \xi $. For $ T \le 30 $\,K, an exponent $ p = 0.66 $ (close to 0.5) rather describes the data well, indicating that two-dimensional (2D) pinning, e.g. at the LAGBs, governs $ J_{\rm c} $. This result indicates that the dominant pinning centers or mechanisms change with temperature. As mentioned above, many small columnar grains having a diameter of 30--55\,nm in NdFeAs(O,F) provide sufficient LAGBs parallel or nearly parallel to $ c $-axis, which can work as effective 2D pinning centers at low temperatures ($ T \le 30 $\,K). Aside from those, no point-like pins of size 4--6\,nm (considering the temperature dependence of coherence length) were visible in this superconducting film. Consequently, there should be other features that can serve as strong pinning centers at high temperatures (35--41\,K).\par

\begin{figure}[htbp]
	\includegraphics[width=0.7\textwidth]{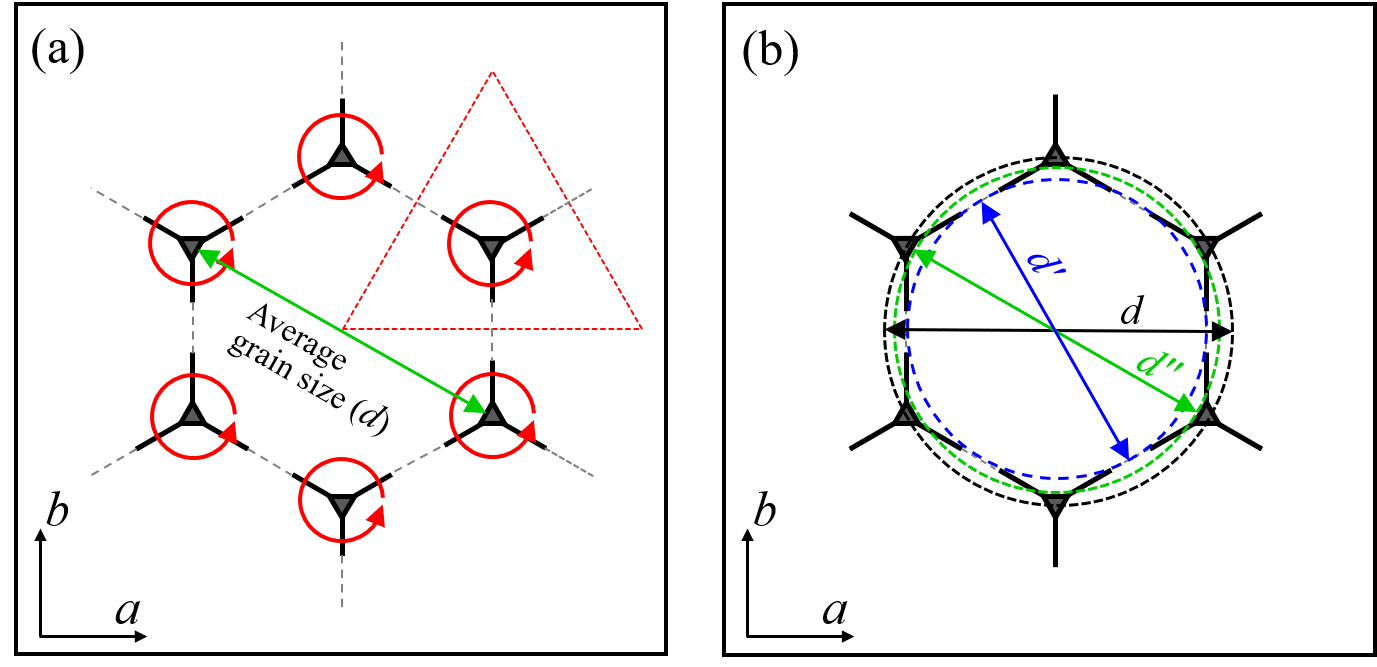}
	\caption{(a) Top view of the hexagonal prismatic grain model with magnetic flux vertices. (b) Three alternative definitions for the diameter of this hexagonal grain: the black diameter $ d $ of circumscribed circle, the blue diameter $ d^{\prime} $ of inscribed circle and the green diameter $ d^{\prime\prime} $ of equivalent area circle. Note that $ \mu_0 H_{\rm m} $ and $ {\mu_0 H_{\rm m}}^{\prime} $ in Equation \ref{eqn:2} are calculated with the hexagonal area, while $ {\mu_0 H_{\rm m}}^{\prime\prime} $ with the circular area.}
	\label{fgr:6}
\end{figure}

After further investigation, the triple junctions (TJs) of the LAGBs were expected to act as effective pinning centers with a larger transverse dimension for fields of $ H \parallel c $, which provide stronger pinning forces than LAGBs.\cite{DewHughes1974} In the following, this hypothesis will be checked. According to the orientation analysis of the film (Figure \ref{fgr:2}), the LAGBs of the epitaxial columnar grains are in general randomly distributed, like a low-angle grain boundary network. It has been shown that the edge number probability of grains in sufficiently large netlike 2D systems sharply peaks around 6, i.e. the most likely grain form is a hexagon.\cite{Kawasaki1989,Smith2015} Due to the columnar grain growth in the NdFeAs(O,F) film on MgO template (Figure \ref{fgr:1},\ref{fgr:2},S2), it is therefore reasonable to model its microstructure solely with hexagonal prismatic grains without loss of generality. In fact, the quasi-hexagonal grain was observed in high-resolution plan-view image of the MgO template (Figure S4). This justifies the validity of our hypothesis, since the texture of MgO template transfers to NdFeAs(O,F). Moreover, a similar structure has been artificially fabricated and analyzed in ref.\cite{Piraux2012}. As illustrated in Figure \ref{fgr:6}a, the mean area per TJ (the red dashed triangle) will be $ 3 \times S / 6 = S / 2 $ ($ S $ is the mean grain area). If the flux lines\cite{Kleiner1964} are matched with TJs one by one (depicted by the red circular arrows in Figure \ref{fgr:6}a), the matching field relational formula will be $ \mathit{\Phi}_0 = \mu_0 H_{\rm m} \times S / 2 $ ($ \mathit{\Phi}_0 = 2.07 \times 10^{-15} $\,Wb is the elementary flux quantum). Then, when substituting the mean value of $ d_{\rm g} $ for the mean grain diameter, for the three alternative definitions of the hexagonal grain diameter shown in Figure \ref{fgr:6}b, the corresponding matching fields are calculated as
\begin{equation}
  \left\{
    \begin{aligned}
      d = 61\,{\rm nm} &\Longrightarrow \mu_0 H_{\rm m} = \frac{2 \mathit{\Phi}_0}{S} = \frac{16 \mathit{\Phi}_0}{3 \sqrt{3} d^2} = 1.7\,{\rm T} \\ 
      d^{\prime} = 61\,{\rm nm} &\Longrightarrow {\mu_0 H_{\rm m}} ^{\prime} = \frac{2 \mathit{\Phi}_0}{S^{\prime}} = \frac{4 \mathit{\Phi}_0}{\sqrt{3} {d^{\prime}}^2} = 1.3\,{\rm T} \\ 
      d^{\prime\prime} = 61\,{\rm nm} &\Longrightarrow {\mu_0 H_{\rm m}} ^{\prime\prime} = \frac{2 \mathit{\Phi}_0}{S^{\prime\prime}} = \frac{8 \mathit{\Phi}_0}{\pi {d^{\prime\prime}}^2} = 1.4\,{\rm T} 
    \end{aligned}
  \right\} 
  \label{eqn:2}
\end{equation} 
i.e. the matching field $ \mu_0 H_{\rm m} = 1.5 \pm 0.2 $\,T. The same field can also be recognized in $ F_{\rm p} $-$ H $ at low temperature as maximum field of the nearly linear behavior (pointed by the arrow in Figure \ref{fgr:5}b and Figure S3c), usually marking single vortex pinning. This gives evidence of the stronger pinning effect of triple junctions, even at high temperatures (35--41\,K) where the LAGBs and small point-like defects cannot contribute anymore. This leads to a possible explanation for the different $ p $ values of $ f_{\rm p} (h) $: At high temperatures, where the field range up to $ \sim \mu_0 H_{\rm m} $ is relatively larger compared to the irreversibility field, vortex pinning is dominated by the TJs. At low temperatures, others, most likely smaller defects such as the LAGBs themselves with occasional dislocations as well as atomic defects, have to come into play because the irreversibility field is now much larger than $ \mu_0 H_{\rm m} $. Considering that the TJs most likely have a larger transverse dimension than the adjacent LAGBs, this explanation is supported by the results of Paturi et al. on YBCO nanocomposite films with $ H \parallel c $.\cite{Paturi2016} They found that the exponent $ p $ increases roughly linearly from $ \sim 0.5 $ for defects in the range of $ \xi $ value towards $ \sim 0.9 $ for the largest defects at the verge of multi-vortex pinning. The $ p $ values of 0.66 and 0.89 of our film lay well in this range. \par

\subsection{Angular maps of in-plane and out-of-plane crystal rotation}

\begin{figure}[htb]
	\includegraphics[width=\textwidth]{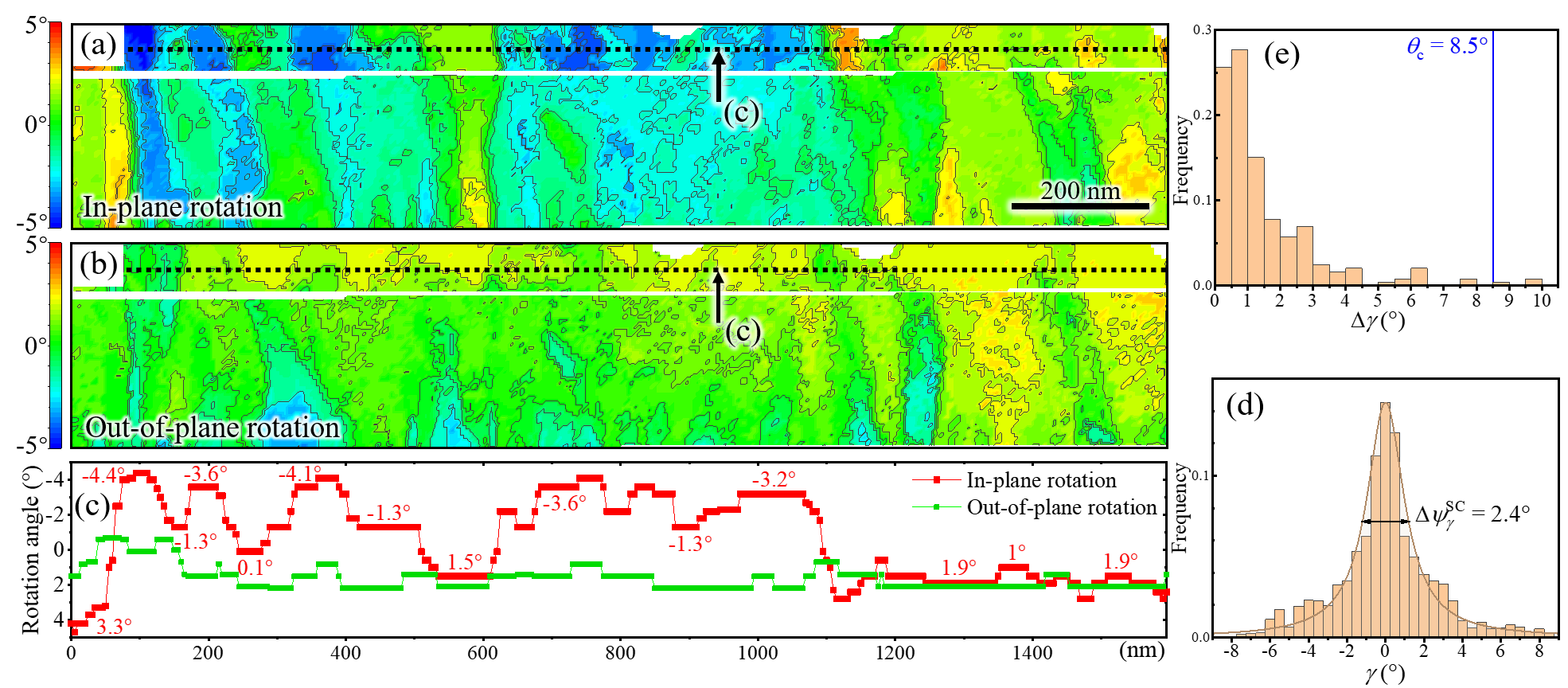}
	\caption{(a) and (b) Angular distribution maps of crystal rotation around $ c $-axis (in-plane crystal rotation) and around a line within $ ab $-plane (out-of-plane crystal rotation), respectively, evaluated from the crystal orientation mapping data in Figure \ref{fgr:2} (detailed information in SI). The difference between adjacent contour lines is $ 1^{\rm \circ} $. (c) The profile along the horizontal dotted lines denoted in (a) and (b). (d) The histogram of the in-plane rotation angular distribution counting all 10 sets of orientation data from the 10 TEM samples, and the relevant Lorentz distribution curve (the best fitting here). The FWHM of this curve is regarded as the statistical in-plane misorientation angle for this NdFeAs(O,F) superconducting (SC) film, $ \Delta \psi_\gamma^{\rm SC} $. (e) The histogram of in-plane misorientation angles between adjacent grains, $ \Delta \gamma $. The critical angle $ \theta_{\rm c} $ regarding the [001]-tilt is marked for comparison.}
	\label{fgr:7}
\end{figure}

In order to thoroughly investigate in-plane and out-of-plane misorientations separately, crystal rotation angles around the $ c $-axis (in-plane rotation angle, $ \gamma $), and around a line within the $ ab $-plane (out-of-plane rotation angle, $ \beta $) of each data point were extracted from the Euler angles of crystal orientation mapping data (Figure \ref{fgr:2}) of the NdFeAs(O,F) film and the MgO template. The detailed extracting method is described in SI (Figure S5) and ref.\cite{Bunge1982,Engler2009}. The resultant in-plane and out-of-plane crystal rotation angular maps are shown in Figure \ref{fgr:7}a,b. The difference between adjacent contour lines is $ 1^{\rm \circ} $. From the clear color contrast and contour lines in the maps, it is evident that the film crystals' size and orientation, both in-plane and out-of-plane, are dominated by those of the MgO template, where the orientation distribution slightly broadens in both cases (see Figure S5 in SI). In addition, it is also visualized that the in-plane misorientation is larger than the out-of-plane misorientation in most domains of NdFeAs(O,F) film and MgO template. \par

Figure \ref{fgr:7}c shows the angular profiles along the horizontal dotted lines in the NdFeAs(O,F) film indicated in Figure \ref{fgr:7}a,b. The angular profile of the in-plane rotation is similar to the misorientation angle profile in Figure \ref{fgr:2}d, while the out-of-plane rotation profile shows a much smaller fluctuation. This feature illustrates again the dominance of the in-plane rotation in the total crystal rotations of this sample. The histogram of the in-plane rotation angle ($ \gamma $) distribution from all 10 data sets, Figure \ref{fgr:7}d, is well described by a Lorentz distribution. This enables us to compare the FWHM of this in-plane rotation angle distribution counted from SPED orientation data ($ \Delta \psi_\gamma^{\rm SC}  = 2.4^{\rm \circ} $ in Figure \ref{fgr:7}d) with the average FWHM of $ \phi $-scan peaks of XRD of this sample ($ \Delta \phi  = 4.0^{\rm \circ} $ in Figure \ref{fgr:1}e). The reason of $ \Delta\phi > \Delta \psi_\gamma^{\rm SC} $ is partially ascribed to the fact that the XRD $ \phi $-scan peaks of in-plane reflections also contain contributions of the out-of-plane crystal rotation.\cite{Specht1998} Furthermore, orientation variations on a larger length scale than measurable by TEM are likely, which contribute to the integral determination in XRD. Besides, the in-plane and out-of-plane rotation angle distributions of the MgO template were also investigated in detail, and can be found in Figure S6. \par

The in-plane misorientation angles between adjacent grains, $ \Delta \gamma $, were also calculated and statistically analyzed, as shown in Figure \ref{fgr:7}e. 99\% of all $ \Delta \gamma $ values are smaller than the critical angle regarding the [001]-tilt grain boundary $ \theta_{\rm c} $ (blue line). \par

\subsection{Microstructural analysis at boundaries and defects}

\begin{figure}[htbp]
	\includegraphics[width=0.9\textwidth]{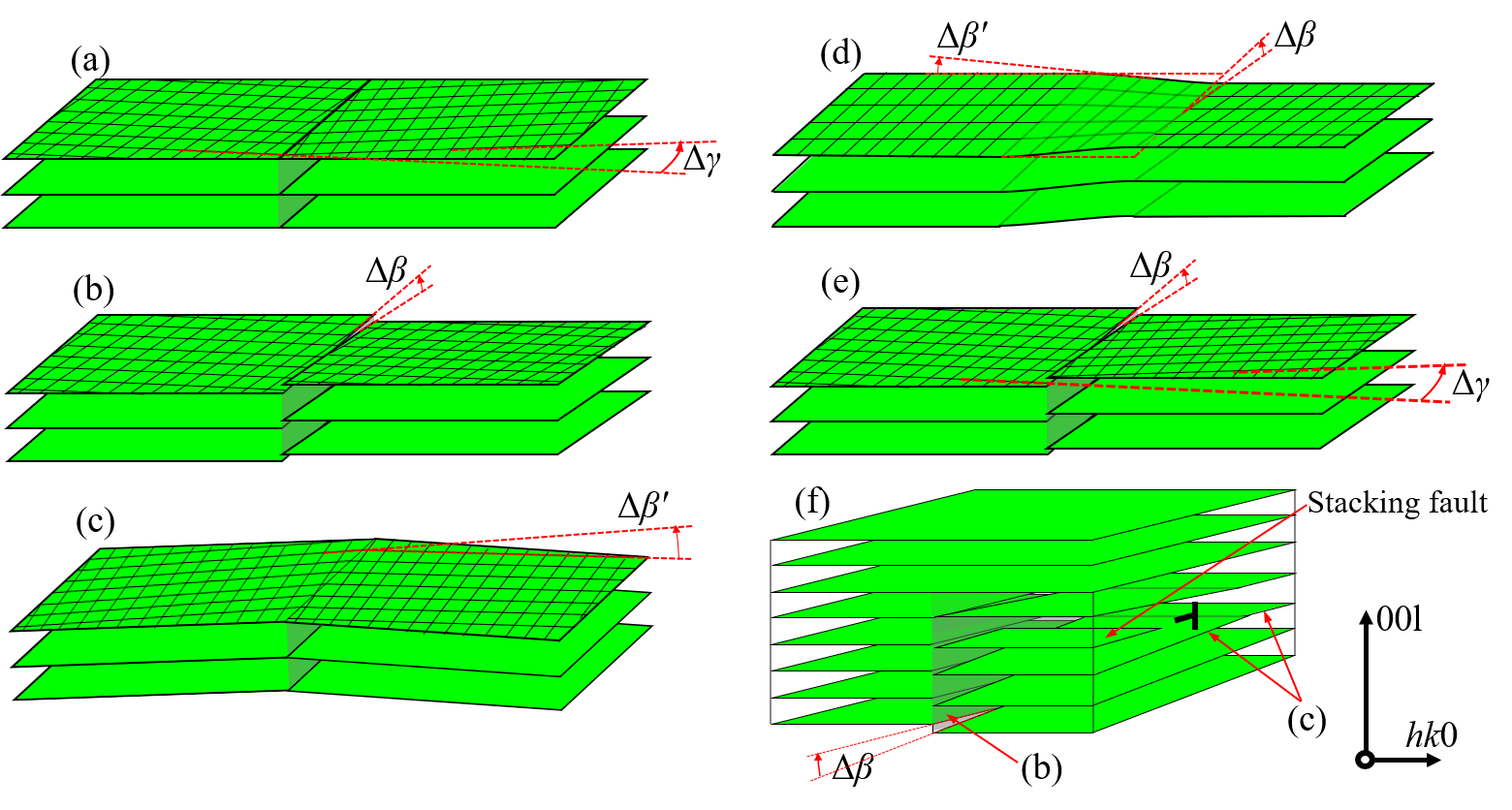}
	\caption{Schematic diagrams of the representative grain boundary regions. (a) The [001]-tilt grain boundary with sole in-plane misorientation ($ \Delta\gamma $). (b), (c) and (d) Three types of out-of-plane misorientation situations ($ \Delta\beta $): the [$ hk0 $]-twist boundary, the [$ hk0 $]-tilt boundary and the mixture of them, respectively. (e) One of the composite grain boundaries with both in-plane and out-of-plane misorientations. (f) A stacking fault compensates the out-of-plane misorientation at the grain boundary.}
	\label{fgr:8}
\end{figure}

In addition, although the in-plane misorientations are the main component, the small yet present out-of-plane misorientations were also observed. The corresponding histogram to Figure \ref{fgr:7}e for out-of-plane misorientation angles, $ \Delta \beta $, is displayed in Figure S7 of SI. It was found in the layered NdFeAs(O,F) structure that the vertical grain boundaries present three different sorts of conformation: the sole in-plane misorientation ($ 0 < \Delta \gamma < \theta_{\rm c} $, $ \Delta \beta = 0 $), the sole out-of-plane misorientation ($ \Delta \gamma = 0 $, $ \Delta \beta > 0 $) and a composite structure of these two states ($ 0 < \Delta \gamma < \theta_{\rm c} $, $ \Delta \beta > 0 $). Some representative diagrams of these three sorts of conformation are sketched in Figure \ref{fgr:8}. As depicted in Figure \ref{fgr:8}a, the [001]-tilt grain boundaries with only in-plane misorientation ($ 0 < \Delta\gamma < \theta_{\rm c} $, $ \Delta \beta = 0 $) can be formed without discontinuing the atomic layers of the conductive FeAs and the charge-reservoir Nd(O,F). On the other hand, the grain boundaries with out-of-plane misorientation ($ \Delta\beta > 0 $) display three types of configurations: the [$ hk0 $]-twist boundary (Figure \ref{fgr:8}b,e), the [$ hk0 $]-tilt boundary (Figure \ref{fgr:8}c) and the mixed state of them (Figure \ref{fgr:8}d). In particular, the out-of-plane misorientations in Figure \ref{fgr:8}b,e will cause disconnections of FeAs and Nd(O,F) atomic layers. Accordingly, it may be initially considered that out-of-plane misorientations have a more severe impact on $ J_{\rm c} $ than in-plane misorientations. However, even in the presence of out-of-plane misorientations, the FeAs and Nd(O,F) atomic layers can maintain or restore their continuity at grain boundaries by continuously curving and/or introducing stacking faults, as the structure depicted in Figure \ref{fgr:8}d,f. \par

\begin{figure}[htbp]
	\includegraphics[width=0.9\textwidth]{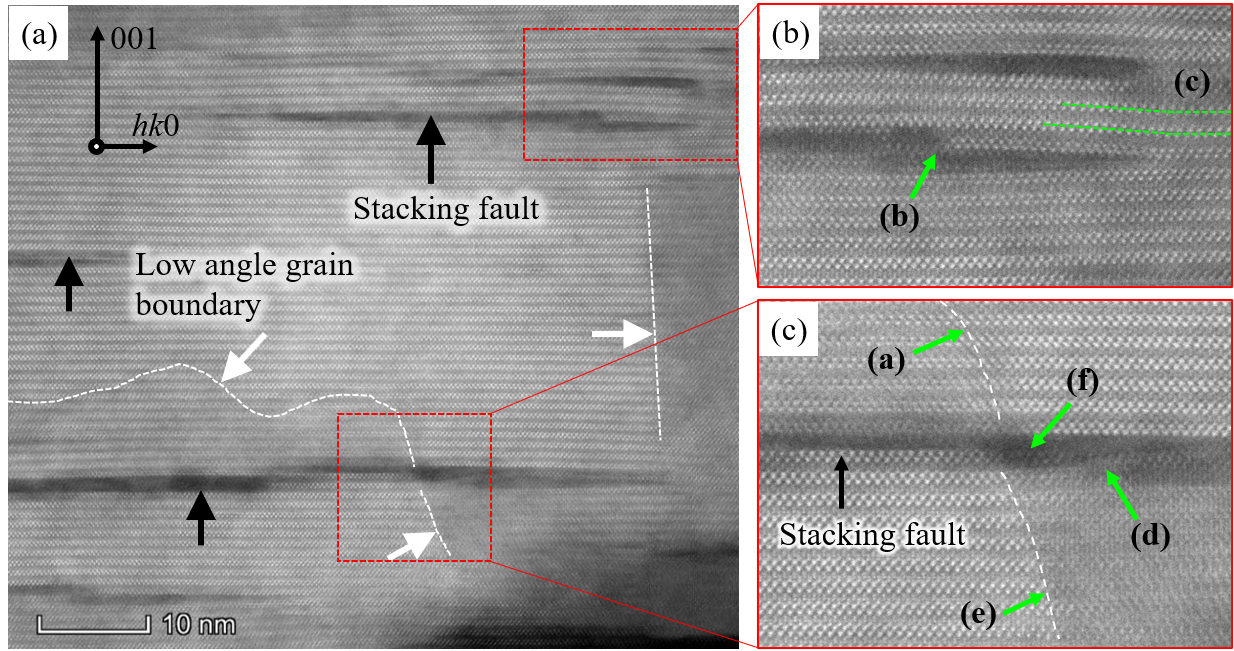}
	\caption{The low-magnification atomic-resolution STEM image of the textured NdFeAs(O,F) thin film (a), and the partially enlarged views (b) and (c). Stacking faults are marked by black arrows. Low angle grain boundaries are marked by white arrows. The green arrows in (b) and (c) point out the microstructures corresponding to various situations in Figure \ref{fgr:8}.}
	\label{fgr:9}
\end{figure}

All configurations exhibited in Figure \ref{fgr:8} have been observed and pointed out one by one in Figure \ref{fgr:9} showing the microstructure on the atomic scale. Moreover, microstructural observations of multiple areas in the ten specimens revealed that the phenomenon of restoring the continuity of atomic layers is universal in this epitaxial NdFeAs(O,F) film. This may be the reason why such a small out-of-plane misorientation did not degrade the $ J_{\rm c} $ of this film apparently. Furthermore, as with the in-plane misorientation, there should also be a critical angle for the out-of-plane misorientation, i.e. $ \Delta\beta_{\rm c} $. When $ \Delta\beta > \Delta\beta_{\rm c} $, most of $ ab $ atomic layers will no longer be able to restore the continuity in grain boundaries. As a result, the dense non-superconducting barriers make the $ J_{\rm c,GB} $ decrease exponentially with $ \Delta\beta $.  \par

\section{Conclusions}

A textured NdFeAs(O,F) thin film grown by MBE on IBAD-MgO substrate exhibited superior superconducting properties: a critical temperature $ T_{\rm c}  = 46 $\,K, self-field $ J_{\rm c} $ up to $ 2 \times 10^6 \,{\rm A/cm^2} $ at 4.2\,K, and in-field $ J_{\rm c} $ for $ H \parallel c $ of $ 10^5 \,{\rm A/cm^2} $ at 10\,T, 15\,K. The nanoscale crystal orientation mapping using SPED verified the highly textured columnar grain structure, where 99\% of in-plane misorientations between adjacent grains were less than $ \theta_{\rm c} $ of $ 8.5^{\rm \circ} $. This demonstrates that the present sample satisfies one of the necessary conditions for high $ J_{\rm c} $ superconducting films. In addition, in the presence of the small out-of-plane misorientations in this epitaxial NdFeAs(O,F) film, the $ ab $ atomic layers can maintain or restore their continuity at grain boundaries by continuously curving and/or introducing stacking faults. As a result, this study supports the effectiveness of grain boundary engineering within $ \theta_{\rm c} $ for the epitaxial Fe-based superconducting films for high field applications. In addition, the triple junctions of grain boundaries are expected to act as pinning centers, even at high temperatures (35--41\,K).

\section{Experimental Methods}

\subsection{Growth of Films and Measurement of Superconducting Properties}

The NdFeAs(O,F) thin films were grown on a $ 10 \times 10 \,{\rm mm^2} $ IBAD-MgO/\ch{Y2O3}/Hastelloy substrate by MBE, as described in detail in ref.\cite{Iida2014}. To obtain superconducting films, a NdOF over-layer was deposited on the mother compound (NdFeAsO) for F doping by diffusion.\cite{Kawaguchi2011} The thickness of \ch{Y2O3}, MgO and NdFeAsO were 710\,nm, 210\,nm and 70\,nm, respectively, confirmed by TEM. After the film growth, the sample was laser-cut into small pieces for structural analysis and electromagnetic property measurements. The temperature dependence of magnetization was measured by a superconducting quantum interference device (SQUID) magnetometer to determine the superconducting transition temperature $ T_{\rm c} $ under a magnetic field $ \mu_0 H $ of 0.8\,mT parallel to the film's $ ab $-plane, $ H \parallel ab $. A small bridge of 0.03\,mm width and 0.5\,mm length was fabricated by laser cutting for electrical transport measurements. To draw the magnetic phase diagram for fields perpendicular to the substrate surface ($ H \parallel c $), the resistivity was measured up to 14\,T by a standard four-probe method (bias current $ I_{\rm b} = 10 $\,\textmu A) using a physical property measurement system (PPMS). The upper critical field ($ H_{\rm c2} $) was evaluated by a temperature-dependent resistivity criterion (90\% of the normal state resistivity). The irreversibility field ($ H_{\rm irr} $) was determined by a constant resistivity criterion, $ \rho _{\rm c} = E_{\rm c} / J_{\rm b} $, where $ E_{\rm c} = 1 $\,\textmu V/cm is the electric field criterion for transport $ J_{\rm c} $ and $ J_{\rm b} = 476\,{\rm A/cm^2} $ is the bias current density, which was also used as criterion for $ H_{\rm irr} $ from $ J_{\rm c} $-$ H $ characteristics. \par

\subsection{Characterization of Nanoscale Texture and Microstructure}

The global texture and the crystal structure were determined by X-ray diffraction (XRD) with Cu-K$ \alpha $ radiation. For analyzing crystal structure and nanoscale crystal orientations by TEM, cross-sectional specimens were prepared by focused ion beam (FIB) milling. The atomic-resolution structural observation was conducted on a Titan Cubed G2 (Thermo Fisher Scientific) under a scanning transmission electron microscopy (STEM) mode with an acceleration voltage of 300\,kV. The annular dark-field (ADF) detection angle range was set from 68\,mrad to 200\,mrad. \par

Nanoscale crystal orientations were evaluated using a transmission electron microscope, Tecnai G2 F20 (Thermo Fisher Scientific) equipped with NanoMEGAS ASTAR$^{\rm TM}$ system at an acceleration voltage of 200\,kV. This combined system enables the mapping of precession electron diffraction (PED) with less than 10\,nm spatial resolution, which yields more detailed information about the orientation distribution and the relationship between epitaxial films and their substrates. Details of this ACOM (or SPED) technique are described in SI and refs.\cite{Rauch2010,Wu2009,Barnard2017}. A noticeable disadvantage of high spatial resolution in TEM is too small a maximum field of view coverable in a single data acquisition. In order to enhance the statistical reliability of this nanoscale crystal-orientation analysis, ten cross-sectional specimens randomly sampled from this superconducting film were investigated, which yielded 247 investigated grains in total whose size and crystal orientation were successfully determined. \par

\begin{acknowledgement}

The authors thank Prof. H. Saito (Kyushu University, Japan) and Dr. C. Wang (The ultramicroscopy Research Center, Kyushu University) for their cooperative supports. This work was supported financially by China Scholarship Council (CSC) under No.201806460012; the Japan Society for the Promotion of Science (JSPS)/Ministry of Education, Culture, Sports, Science and Technology (MEXT), Japan KAKENHI (JP18K18954, JP18H05479, JP20H02681, JP20H02426); Japan-German Research Cooperative Program between JSPS and DAAD, Grant number JPJSBP120203506; and Japan Science and Technology Agency (JST) CREST (\#JPMJCR18J4).

\end{acknowledgement}

\begin{suppinfo}

The Supporting Information is available free of charge.
\begin{itemize}
  \item Detailed interpretations of scanning precession electron diffraction (SPED) technique and in-plane / out-of-plane crystal rotation angles; Supplementary data on superconducting properties and crystal orientation of the NdFeAs(O,F) thin film; Additional characterization about the nanoscale texture of the IBAD-MgO substrate; Additional explanations for some confusing symbols in this article.
\end{itemize}

\end{suppinfo}

\bibliography{achemso}

\providecommand{\latin}[1]{#1}
\makeatletter
\providecommand{\doi}
  {\begingroup\let\do\@makeother\dospecials
  \catcode`\{=1 \catcode`\}=2 \doi@aux}
\providecommand{\doi@aux}[1]{\endgroup\texttt{#1}}
\makeatother
\providecommand*\mcitethebibliography{\thebibliography}
\csname @ifundefined\endcsname{endmcitethebibliography}
  {\let\endmcitethebibliography\endthebibliography}{}
\begin{mcitethebibliography}{34}
\providecommand*\natexlab[1]{#1}
\providecommand*\mciteSetBstSublistMode[1]{}
\providecommand*\mciteSetBstMaxWidthForm[2]{}
\providecommand*\mciteBstWouldAddEndPuncttrue
  {\def\EndOfBibitem{\unskip.}}
\providecommand*\mciteBstWouldAddEndPunctfalse
  {\let\EndOfBibitem\relax}
\providecommand*\mciteSetBstMidEndSepPunct[3]{}
\providecommand*\mciteSetBstSublistLabelBeginEnd[3]{}
\providecommand*\EndOfBibitem{}
\mciteSetBstSublistMode{f}
\mciteSetBstMaxWidthForm{subitem}{(\alph{mcitesubitemcount})}
\mciteSetBstSublistLabelBeginEnd
  {\mcitemaxwidthsubitemform\space}
  {\relax}
  {\relax}

\bibitem[Kamihara \latin{et~al.}(2008)Kamihara, Watanabe, Hirano, and
  Hosono]{Kamihara2008}
Kamihara,~Y.; Watanabe,~T.; Hirano,~M.; Hosono,~H. {Iron-based layered
  superconductor La[O1-xFx]FeAs (x= 0.05-0.12) with Tc = 26 K}. \emph{J. Am.
  Chem. Soc.} \textbf{2008}, \emph{130}, 3296--3297\relax
\mciteBstWouldAddEndPuncttrue
\mciteSetBstMidEndSepPunct{\mcitedefaultmidpunct}
{\mcitedefaultendpunct}{\mcitedefaultseppunct}\relax
\EndOfBibitem
\bibitem[Fujioka \latin{et~al.}(2013)Fujioka, Denholme, Ozaki, Okazaki,
  Deguchi, Demura, Hara, Watanabe, Takeya, Yamaguchi, Kumakura, and
  Takano]{Fujioka2013}
Fujioka,~M.; Denholme,~S.~J.; Ozaki,~T.; Okazaki,~H.; Deguchi,~K.; Demura,~S.;
  Hara,~H.; Watanabe,~T.; Takeya,~H.; Yamaguchi,~T.; Kumakura,~H.; Takano,~Y.
  {Phase diagram and superconductivity at 58.1 K in $\alpha$-FeAs-free
  SmFeAsO1-xFx}. \emph{Supercond. Sci. Technol.} \textbf{2013}, \emph{26},
  085023\relax
\mciteBstWouldAddEndPuncttrue
\mciteSetBstMidEndSepPunct{\mcitedefaultmidpunct}
{\mcitedefaultendpunct}{\mcitedefaultseppunct}\relax
\EndOfBibitem
\bibitem[Pallecchi \latin{et~al.}(2015)Pallecchi, Eisterer, Malagoli, and
  Putti]{Pallecchi2015}
Pallecchi,~I.; Eisterer,~M.; Malagoli,~A.; Putti,~M. {Application potential of
  Fe-based superconductors}. \emph{Supercond. Sci. Technol.} \textbf{2015},
  \emph{28}, 114005\relax
\mciteBstWouldAddEndPuncttrue
\mciteSetBstMidEndSepPunct{\mcitedefaultmidpunct}
{\mcitedefaultendpunct}{\mcitedefaultseppunct}\relax
\EndOfBibitem
\bibitem[H{\"{a}}nisch \latin{et~al.}(2019)H{\"{a}}nisch, Iida, H{\"{u}}hne,
  and Tarantini]{Haenisch2019}
H{\"{a}}nisch,~J.; Iida,~K.; H{\"{u}}hne,~R.; Tarantini,~C. {Fe-based
  superconducting thin films-preparation and tuning of superconducting
  properties}. \emph{Supercond. Sci. Technol.} \textbf{2019}, \emph{32},
  093001\relax
\mciteBstWouldAddEndPuncttrue
\mciteSetBstMidEndSepPunct{\mcitedefaultmidpunct}
{\mcitedefaultendpunct}{\mcitedefaultseppunct}\relax
\EndOfBibitem
\bibitem[Durrell \latin{et~al.}(2011)Durrell, Eom, Gurevich, Hellstrom,
  Tarantini, Yamamoto, and Larbalestier]{Durrell2011}
Durrell,~J.~H.; Eom,~C.~B.; Gurevich,~A.; Hellstrom,~E.~E.; Tarantini,~C.;
  Yamamoto,~A.; Larbalestier,~D.~C. {The behavior of grain boundaries in the
  Fe-based superconductors}. \emph{Reports Prog. Phys.} \textbf{2011},
  \emph{74}, 124511\relax
\mciteBstWouldAddEndPuncttrue
\mciteSetBstMidEndSepPunct{\mcitedefaultmidpunct}
{\mcitedefaultendpunct}{\mcitedefaultseppunct}\relax
\EndOfBibitem
\bibitem[Iida \latin{et~al.}(2020)Iida, H{\"{a}}nisch, and Yamamoto]{Iida2020}
Iida,~K.; H{\"{a}}nisch,~J.; Yamamoto,~A. {Grain boundary characteristics of
  Fe-based superconductors}. \emph{Supercond. Sci. Technol.} \textbf{2020},
  \emph{33}, 043001\relax
\mciteBstWouldAddEndPuncttrue
\mciteSetBstMidEndSepPunct{\mcitedefaultmidpunct}
{\mcitedefaultendpunct}{\mcitedefaultseppunct}\relax
\EndOfBibitem
\bibitem[Iida \latin{et~al.}(2019)Iida, Omura, Matsumoto, Hatano, and
  Ikuta]{Iida2019}
Iida,~K.; Omura,~T.; Matsumoto,~T.; Hatano,~T.; Ikuta,~H. {Grain boundary
  characteristics of oxypnictide NdFeAs(O,F) superconductors}. \emph{Supercond.
  Sci. Technol.} \textbf{2019}, \emph{32}, 074003\relax
\mciteBstWouldAddEndPuncttrue
\mciteSetBstMidEndSepPunct{\mcitedefaultmidpunct}
{\mcitedefaultendpunct}{\mcitedefaultseppunct}\relax
\EndOfBibitem
\bibitem[Katase \latin{et~al.}(2011)Katase, Ishimaru, Tsukamoto, Hiramatsu,
  Kamiya, Tanabe, and Hosono]{Katase2011}
Katase,~T.; Ishimaru,~Y.; Tsukamoto,~A.; Hiramatsu,~H.; Kamiya,~T.; Tanabe,~K.;
  Hosono,~H. {Advantageous grain boundaries in iron pnictide superconductors}.
  \emph{Nat. Commun.} \textbf{2011}, \emph{2}, 409\relax
\mciteBstWouldAddEndPuncttrue
\mciteSetBstMidEndSepPunct{\mcitedefaultmidpunct}
{\mcitedefaultendpunct}{\mcitedefaultseppunct}\relax
\EndOfBibitem
\bibitem[Si \latin{et~al.}(2015)Si, Zhang, Shi, Ozaki, Jaroszynski, and
  Li]{Si_2015}
Si,~W.; Zhang,~C.; Shi,~X.; Ozaki,~T.; Jaroszynski,~J.; Li,~Q. {Grain boundary
  junctions of FeSe0.5Te0.5 thin films on SrTiO3 bi-crystal substrates}.
  \emph{Appl. Phys. Lett.} \textbf{2015}, \emph{106}, 032602\relax
\mciteBstWouldAddEndPuncttrue
\mciteSetBstMidEndSepPunct{\mcitedefaultmidpunct}
{\mcitedefaultendpunct}{\mcitedefaultseppunct}\relax
\EndOfBibitem
\bibitem[Sarnelli \latin{et~al.}(2017)Sarnelli, Nappi, Camerlingo, Enrico,
  Bellingeri, Kawale, Braccini, Leveratto, and Ferdeghini]{Sarnelli2017}
Sarnelli,~E.; Nappi,~C.; Camerlingo,~C.; Enrico,~E.; Bellingeri,~E.;
  Kawale,~S.; Braccini,~V.; Leveratto,~A.; Ferdeghini,~C. {Properties of Fe(Se,
  Te) Bicrystal Grain Boundary Junctions, SQUIDs, and Nanostrips}. \emph{IEEE
  Trans. Appl. Supercond.} \textbf{2017}, \emph{27}, 7400104\relax
\mciteBstWouldAddEndPuncttrue
\mciteSetBstMidEndSepPunct{\mcitedefaultmidpunct}
{\mcitedefaultendpunct}{\mcitedefaultseppunct}\relax
\EndOfBibitem
\bibitem[Sato \latin{et~al.}(2016)Sato, Hiramatsu, Kamiya, and
  Hosono]{Sato2016}
Sato,~H.; Hiramatsu,~H.; Kamiya,~T.; Hosono,~H. {Enhanced critical-current in
  P-doped BaFe2As2 thin films on metal substrates arising from poorly aligned
  grain boundaries}. \emph{Sci. Rep.} \textbf{2016}, \emph{6}, 36828\relax
\mciteBstWouldAddEndPuncttrue
\mciteSetBstMidEndSepPunct{\mcitedefaultmidpunct}
{\mcitedefaultendpunct}{\mcitedefaultseppunct}\relax
\EndOfBibitem
\bibitem[Kitaguchi \latin{et~al.}(2004)Kitaguchi, Matsumoto, Kumakura, Doi,
  Yamamoto, Saitoh, Sosiati, and Hata]{Kitaguchi2004}
Kitaguchi,~H.; Matsumoto,~A.; Kumakura,~H.; Doi,~T.; Yamamoto,~H.; Saitoh,~K.;
  Sosiati,~H.; Hata,~S. {MgB2 films with very high critical current densities
  due to strong grain boundary pinning}. \emph{Appl. Phys. Lett.}
  \textbf{2004}, \emph{85}, 2842--2844\relax
\mciteBstWouldAddEndPuncttrue
\mciteSetBstMidEndSepPunct{\mcitedefaultmidpunct}
{\mcitedefaultendpunct}{\mcitedefaultseppunct}\relax
\EndOfBibitem
\bibitem[Iida \latin{et~al.}(2014)Iida, Kurth, Chihara, Sumiya, Grinenko,
  Ichinose, Tsukada, H{\"{a}}nisch, Matias, Hatano, Holzapfel, and
  Ikuta]{Iida2014}
Iida,~K.; Kurth,~F.; Chihara,~M.; Sumiya,~N.; Grinenko,~V.; Ichinose,~A.;
  Tsukada,~I.; H{\"{a}}nisch,~J.; Matias,~V.; Hatano,~T.; Holzapfel,~B.;
  Ikuta,~H. {Highly textured oxypnictide superconducting thin films on metal
  substrates}. \emph{Appl. Phys. Lett.} \textbf{2014}, \emph{105}, 172602\relax
\mciteBstWouldAddEndPuncttrue
\mciteSetBstMidEndSepPunct{\mcitedefaultmidpunct}
{\mcitedefaultendpunct}{\mcitedefaultseppunct}\relax
\EndOfBibitem
\bibitem[Sheehan \latin{et~al.}(2011)Sheehan, Jung, Holesinger, Feldmann,
  Edney, Ihlefeld, Clem, and Matias]{Sheehan2011}
Sheehan,~C.; Jung,~Y.; Holesinger,~T.; Feldmann,~D.~M.; Edney,~C.;
  Ihlefeld,~J.~F.; Clem,~P.~G.; Matias,~V. {Solution deposition planarization
  of long-length flexible substrates}. \emph{Appl. Phys. Lett.} \textbf{2011},
  \emph{98}, 071907\relax
\mciteBstWouldAddEndPuncttrue
\mciteSetBstMidEndSepPunct{\mcitedefaultmidpunct}
{\mcitedefaultendpunct}{\mcitedefaultseppunct}\relax
\EndOfBibitem
\bibitem[Iijima \latin{et~al.}(1992)Iijima, Tanabe, Kohno, and
  Ikeno]{Iijima1992}
Iijima,~Y.; Tanabe,~N.; Kohno,~O.; Ikeno,~Y. {In-plane aligned YBa2Cu3O7-x thin
  films deposited on polycrystalline metallic substrates}. \emph{Appl. Phys.
  Lett.} \textbf{1992}, \emph{60}, 769--771\relax
\mciteBstWouldAddEndPuncttrue
\mciteSetBstMidEndSepPunct{\mcitedefaultmidpunct}
{\mcitedefaultendpunct}{\mcitedefaultseppunct}\relax
\EndOfBibitem
\bibitem[Kito \latin{et~al.}(2008)Kito, Eisaki, and Iyo]{Kito2008}
Kito,~H.; Eisaki,~H.; Iyo,~A. {Superconductivity at 54 K in F-Free NdFeAsO 1-
  y}. \emph{J. Phys. Soc. Japan} \textbf{2008}, \emph{77}, 063707\relax
\mciteBstWouldAddEndPuncttrue
\mciteSetBstMidEndSepPunct{\mcitedefaultmidpunct}
{\mcitedefaultendpunct}{\mcitedefaultseppunct}\relax
\EndOfBibitem
\bibitem[Sasaki \latin{et~al.}(1979)Sasaki, Takeuchi, and Fujino]{SASAKI1979}
Sasaki,~S.; Takeuchi,~Y.; Fujino,~K. {X-Ray Determination of Electron-Density
  Distributions in Oxides, MgO, MnO, CoO, and NiO, and Atomic Scattering
  Factors of their Constituent Atoms}. \emph{Proc. Japan Acad. Ser. B Phys.
  Biol. Sci.} \textbf{1979}, \emph{55}, 43--48\relax
\mciteBstWouldAddEndPuncttrue
\mciteSetBstMidEndSepPunct{\mcitedefaultmidpunct}
{\mcitedefaultendpunct}{\mcitedefaultseppunct}\relax
\EndOfBibitem
\bibitem[H{\"{a}}nisch \latin{et~al.}(2014)H{\"{a}}nisch, Iida, Kurth,
  Thersleff, Trommler, Reich, H{\"{u}}hne, Schultz, and
  Holzapfel]{Haenisch2014}
H{\"{a}}nisch,~J.; Iida,~K.; Kurth,~F.; Thersleff,~T.; Trommler,~S.; Reich,~E.;
  H{\"{u}}hne,~R.; Schultz,~L.; Holzapfel,~B. {The effect of 45° grain
  boundaries and associated Fe particles on Jc and resistivity in
  Ba(Fe0.9Co0.1)2As2 thin films}. AIP Conf. Proc. 2014; pp 260--267\relax
\mciteBstWouldAddEndPuncttrue
\mciteSetBstMidEndSepPunct{\mcitedefaultmidpunct}
{\mcitedefaultendpunct}{\mcitedefaultseppunct}\relax
\EndOfBibitem
\bibitem[Iida \latin{et~al.}(2017)Iida, Sato, Tarantini, H{\"{a}}nisch,
  Jaroszynski, Hiramatsu, Holzapfel, and Hosono]{Iida2017}
Iida,~K.; Sato,~H.; Tarantini,~C.; H{\"{a}}nisch,~J.; Jaroszynski,~J.;
  Hiramatsu,~H.; Holzapfel,~B.; Hosono,~H. {High-field transport properties of
  a P-doped BaFe2As2 film on technical substrate.} \emph{Sci. Rep.}
  \textbf{2017}, \emph{7}, 39951\relax
\mciteBstWouldAddEndPuncttrue
\mciteSetBstMidEndSepPunct{\mcitedefaultmidpunct}
{\mcitedefaultendpunct}{\mcitedefaultseppunct}\relax
\EndOfBibitem
\bibitem[Tarantini \latin{et~al.}(2016)Tarantini, Iida, H{\"{a}}nisch, Kurth,
  Jaroszynski, Sumiya, Chihara, Hatano, Ikuta, Schmidt, Seidel, Holzapfel, and
  Larbalestier]{Tarantini2016}
Tarantini,~C.; Iida,~K.; H{\"{a}}nisch,~J.; Kurth,~F.; Jaroszynski,~J.;
  Sumiya,~N.; Chihara,~M.; Hatano,~T.; Ikuta,~H.; Schmidt,~S.; Seidel,~P.;
  Holzapfel,~B.; Larbalestier,~D.~C. {Intrinsic and extrinsic pinning in
  NdFeAs(O,F): vortex trapping and lock-in by the layered structure}.
  \emph{Sci. Rep.} \textbf{2016}, \emph{6}, 36047\relax
\mciteBstWouldAddEndPuncttrue
\mciteSetBstMidEndSepPunct{\mcitedefaultmidpunct}
{\mcitedefaultendpunct}{\mcitedefaultseppunct}\relax
\EndOfBibitem
\bibitem[Dew-Hughes(1974)]{DewHughes1974}
Dew-Hughes,~D. {Flux pinning mechanisms in type II superconductors}.
  \emph{Philos. Mag.} \textbf{1974}, \emph{30}, 293--305\relax
\mciteBstWouldAddEndPuncttrue
\mciteSetBstMidEndSepPunct{\mcitedefaultmidpunct}
{\mcitedefaultendpunct}{\mcitedefaultseppunct}\relax
\EndOfBibitem
\bibitem[Kawasak \latin{et~al.}(1989)Kawasak, Naga, and
  Nakashima]{Kawasaki1989}
Kawasak,~K.; Naga,~T.; Nakashima,~K. {Vertex models for two-dimensional grain
  growth}. \emph{Philos. Mag. B Phys. Condens. Matter; Stat. Mech. Electron.
  Opt. Magn. Prop.} \textbf{1989}, \emph{60}, 399--421\relax
\mciteBstWouldAddEndPuncttrue
\mciteSetBstMidEndSepPunct{\mcitedefaultmidpunct}
{\mcitedefaultendpunct}{\mcitedefaultseppunct}\relax
\EndOfBibitem
\bibitem[Smith(2015)]{Smith2015}
Smith,~C.~S. {Grain Shapes and Other Metallurgical Applications of Topology}.
  \emph{Metallogr. Microstruct. Anal.} \textbf{2015}, \emph{4}, 543--567\relax
\mciteBstWouldAddEndPuncttrue
\mciteSetBstMidEndSepPunct{\mcitedefaultmidpunct}
{\mcitedefaultendpunct}{\mcitedefaultseppunct}\relax
\EndOfBibitem
\bibitem[Piraux and Hallet(2012)Piraux, and Hallet]{Piraux2012}
Piraux,~L.; Hallet,~X. {Artificial vortex pinning arrays in superconducting
  films deposited on highly ordered anodic alumina templates}.
  \emph{Nanotechnology} \textbf{2012}, \emph{23}, 355301\relax
\mciteBstWouldAddEndPuncttrue
\mciteSetBstMidEndSepPunct{\mcitedefaultmidpunct}
{\mcitedefaultendpunct}{\mcitedefaultseppunct}\relax
\EndOfBibitem
\bibitem[Kleiner \latin{et~al.}(1964)Kleiner, Roth, and Autler]{Kleiner1964}
Kleiner,~W.~H.; Roth,~L.~M.; Autler,~S.~H. {Bulk Solution of Ginzburg-Landau
  Equations for Type II Superconductors: Upper Critical Field Region}.
  \emph{Phys. Rev.} \textbf{1964}, \emph{133}, A1226--A1227\relax
\mciteBstWouldAddEndPuncttrue
\mciteSetBstMidEndSepPunct{\mcitedefaultmidpunct}
{\mcitedefaultendpunct}{\mcitedefaultseppunct}\relax
\EndOfBibitem
\bibitem[Paturi \latin{et~al.}(2016)Paturi, Malmivirta, Palonen, and
  Huhtinen]{Paturi2016}
Paturi,~P.; Malmivirta,~M.; Palonen,~H.; Huhtinen,~H. {Dopant diameter
  dependence of Jc(B) in doped YBCO films}. \emph{IEEE Trans. Appl. Supercond.}
  \textbf{2016}, \emph{26}, 8000705\relax
\mciteBstWouldAddEndPuncttrue
\mciteSetBstMidEndSepPunct{\mcitedefaultmidpunct}
{\mcitedefaultendpunct}{\mcitedefaultseppunct}\relax
\EndOfBibitem
\bibitem[Bunge(1982)]{Bunge1982}
Bunge,~H.-J. \emph{Texture Anal. Mater. Sci.}; Elsevier, 1982; pp 3--41\relax
\mciteBstWouldAddEndPuncttrue
\mciteSetBstMidEndSepPunct{\mcitedefaultmidpunct}
{\mcitedefaultendpunct}{\mcitedefaultseppunct}\relax
\EndOfBibitem
\bibitem[Engler and Randle(2009)Engler, and Randle]{Engler2009}
Engler,~O.; Randle,~V. \emph{Introd. to Texture Anal.}; CRC Press, 2009; pp
  15--50\relax
\mciteBstWouldAddEndPuncttrue
\mciteSetBstMidEndSepPunct{\mcitedefaultmidpunct}
{\mcitedefaultendpunct}{\mcitedefaultseppunct}\relax
\EndOfBibitem
\bibitem[Specht \latin{et~al.}(1998)Specht, Goyal, Lee, List, Kroeger,
  Paranthaman, Williams, and Christen]{Specht1998}
Specht,~E.~D.; Goyal,~A.; Lee,~D.~F.; List,~F.~A.; Kroeger,~D.~M.;
  Paranthaman,~M.; Williams,~R.~K.; Christen,~D.~K. {Cube-textured nickel
  substrates for high-temperature superconductors}. \emph{Supercond. Sci.
  Technol.} \textbf{1998}, \emph{11}, 945--949\relax
\mciteBstWouldAddEndPuncttrue
\mciteSetBstMidEndSepPunct{\mcitedefaultmidpunct}
{\mcitedefaultendpunct}{\mcitedefaultseppunct}\relax
\EndOfBibitem
\bibitem[Kawaguchi \latin{et~al.}(2011)Kawaguchi, Uemura, Ohno, Tabuchi,
  Ujihara, Takeda, and Ikuta]{Kawaguchi2011}
Kawaguchi,~T.; Uemura,~H.; Ohno,~T.; Tabuchi,~M.; Ujihara,~T.; Takeda,~Y.;
  Ikuta,~H. {Molecular Beam Epitaxy Growth of Superconducting NdFeAs(O,F) Thin
  Films Using a F-Getter and a Novel F-Doping Method}. \emph{Appl. Phys.
  Express} \textbf{2011}, \emph{4}, 083102\relax
\mciteBstWouldAddEndPuncttrue
\mciteSetBstMidEndSepPunct{\mcitedefaultmidpunct}
{\mcitedefaultendpunct}{\mcitedefaultseppunct}\relax
\EndOfBibitem
\bibitem[Rauch \latin{et~al.}(2010)Rauch, Portillo, Nicolopoulos, Bultreys,
  Rouvimov, and Moeck]{Rauch2010}
Rauch,~E.~F.; Portillo,~J.; Nicolopoulos,~S.; Bultreys,~D.; Rouvimov,~S.;
  Moeck,~P. {Automated nanocrystal orientation and phase mapping in the
  transmission electron microscope on the basis of precession electron
  diffraction}. \emph{Zeitschrift fur Krist.} \textbf{2010}, \emph{225},
  103--109\relax
\mciteBstWouldAddEndPuncttrue
\mciteSetBstMidEndSepPunct{\mcitedefaultmidpunct}
{\mcitedefaultendpunct}{\mcitedefaultseppunct}\relax
\EndOfBibitem
\bibitem[Wu and Zaefferer(2009)Wu, and Zaefferer]{Wu2009}
Wu,~G.; Zaefferer,~S. {Advances in TEM orientation microscopy by combination of
  dark-field conical scanning and improved image matching}.
  \emph{Ultramicroscopy} \textbf{2009}, \emph{109}, 1317--1325\relax
\mciteBstWouldAddEndPuncttrue
\mciteSetBstMidEndSepPunct{\mcitedefaultmidpunct}
{\mcitedefaultendpunct}{\mcitedefaultseppunct}\relax
\EndOfBibitem
\bibitem[Barnard \latin{et~al.}(2017)Barnard, Johnstone, and
  Midgley]{Barnard2017}
Barnard,~J.~S.; Johnstone,~D.~N.; Midgley,~P.~A. {High-resolution scanning
  precession electron diffraction: Alignment and spatial resolution}.
  \emph{Ultramicroscopy} \textbf{2017}, \emph{174}, 79--88\relax
\mciteBstWouldAddEndPuncttrue
\mciteSetBstMidEndSepPunct{\mcitedefaultmidpunct}
{\mcitedefaultendpunct}{\mcitedefaultseppunct}\relax
\EndOfBibitem
\end{mcitethebibliography}



\section*{Supplementary material}

**Supporting Information**

# Nanoscale Texture and Microstructure in NdFeAs(O,F)/IBAD-MgO Superconducting Thin Film with Superior Critical Current Properties

Zimeng Guo,[†,‡] Hongye Gao,[¶] Keisuke Kondo,[§] Takafumi Hatano,[§,‡] Kazumasa Iida,[§,‡] Jens Hänisch,[∥] Hiroshi Ikuta,[§] and Satoshi Hata[*,†,‡,¶,⊥]

[†] *Department of Applied Science for Electronics and Materials, Kyushu University, Fukuoka 816-8580, Japan*

[‡] *Core Research for Evolutionary Science and Technology, Japan Science and Technology Agency, Tokyo 102-0076, Japan*

[¶] *The Ultramicroscopy Research Center, Kyushu University, Fukuoka 819-0395, Japan*

[§] *Department of Materials Physics, Nagoya University, Nagoya 464-8603, Japan*

[∥] *Institute for Technical Physics, Karlsruhe Institute of Technology, Hermann-von-Helmholtz-Platz 1, 76344 Eggenstein-Leopoldshafen, Germany*

[⊥] *Department of Advanced Materials Science and Engineering, Faculty of Engineering Sciences, Kyushu University, Fukuoka 816-8580, Japan*

*E-mail: hata.satoshi.207@m.kyushu-u.ac.jp

1. **Scanning precession electron diffraction**

Over the last decades, electron back-scatter diffraction (EBSD) has played a significant role for the characterization of polycrystalline bulk materials.[1] However, the spatial resolution of EBSD is limited to several tens of nanometers in general.[2] The scanning precession electron diffraction (SPED) technique performed in TEM has recently been considered as an alternative technique with higher spatial resolution, and more feasible for characterization of nanocrystalline materials.[1–3] Figure S1 shows a schematic drawing of an experimental setup for SPED. The beam control device, produced by NanoMEGAS, was used for precession of the incident electron beam around the optical axis by a precession angle $\phi$, and de-precession of the diffracted beams below the sample. This process of beam precession and de-precession is equivalent to the precession of a TEM foil sample around a stationary electron beam.[4] The beam precession exhibits two significant advantages:[3] (1) the number of diffraction spots increases because the Ewald sphere crosses more reciprocal lattice points; (2) the thickness-dependence of the reflection intensity becomes gentle, in other words, the reflections are more 'kinematic-like'.[5–7] Consequently, the precession electron diffraction (PED) pattern will display more diffraction spots even in non-zone axis and particularly be amenable for structure solution problems.[8–10] Meanwhile, by successively shifting the precession incident beam at a selected nanoscale step size and acquiring the PED pattern of each beam position, a series of PED patterns of the region of interest was produced as a form of 4-dimensional data-set (2D diffraction patterns distributing in 2D detected area). Subsequently, by matching the SPED patterns with pre-generated template patterns, the phase and crystal orientation of the interesting region can be mapped.[3]

The SPED analysis in this report was operated by using a transmission electron microscope, Tecnai G2 F20 (Thermo Fisher Sciecntific). The convergence semi-angle has been calculated about 1 mrad from non-precession electron diffraction patterns, and the spatial resolution, $δ_d$, was calculated 1.5 nm in a non-precession mode. Considering the influence of beam precession and possible pivot point misalignment on resolution, PED patterns were acquired in 5 nm steps. Additionally, the precession angle was set to 0.55° (typical range: 0.5° ~ 3°) to reduce its impact on spatial and angular resolution.[3] The precession frequency was 100 Hz, and PED patterns were captured at 25 fps. These settings mean every frame of the PED pattern covered 4 times a full precession cycle, which guarantees a high quality of the patterns. Figure S2 displays one set of typical orientation maps obtained by SPED with the relevant color codes, which shows the nanoscale crystal orientation distribution in a cross-section TEM foil of the epitaxially grown NdFeAs(O,F) film on IBAD-MgO substrate (the sample for this report).

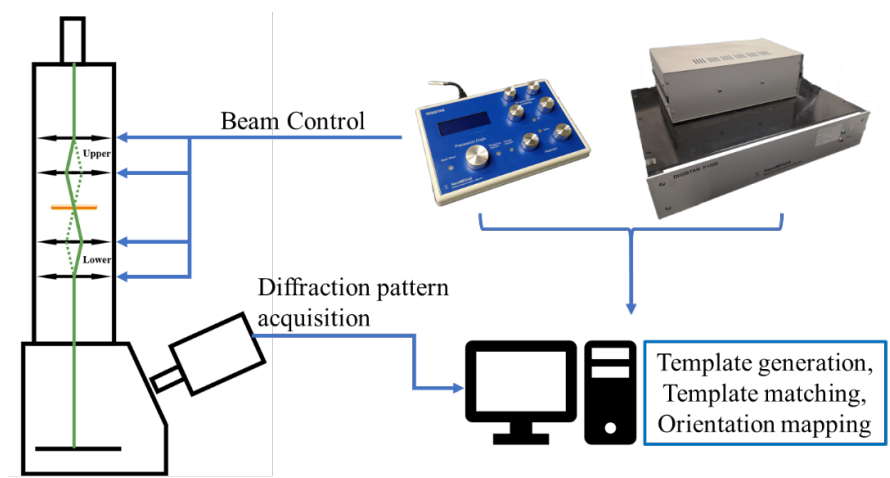

Figure S1. Schematic diagram of the orientation microscopy system for performing SPED analysis.

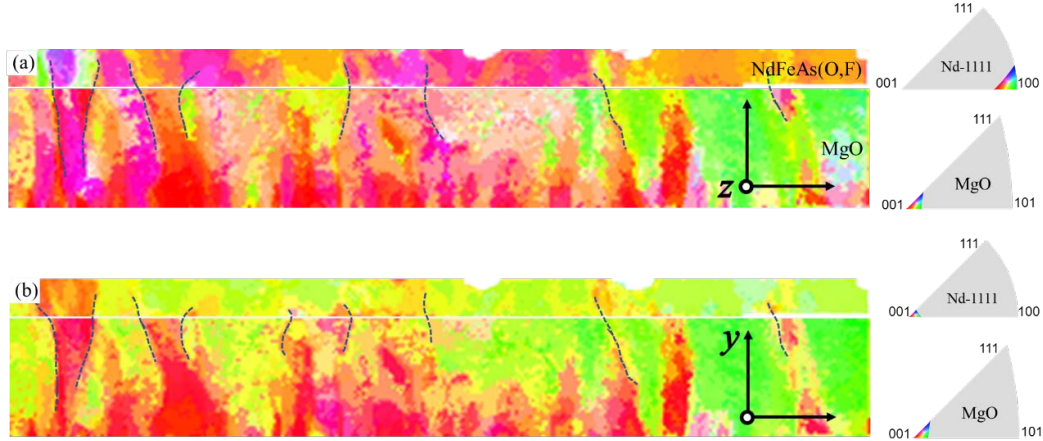

Figure S2. (a) and (b) The orientation maps of the textured NdFeAs(O,F) film epitaxially grown on IBAD-MgO substrate shown on $z$-axis and $y$-axis view direction, respectively, with the corresponding color codes.

## 2. More information about superconducting properties characterization

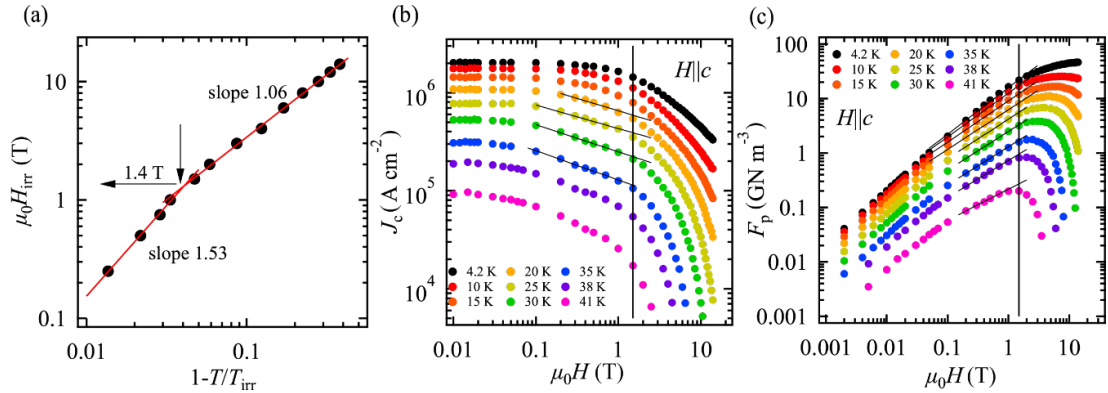

Figure S3. (a) The power law relation of $\mu_0 H_{irr}$ vs. $(1-T/T_{irr})$ in double-logarithmic representation, corresponding to Figure 4d in this report. The intersection of the fitting lines shows clearly the point of change in slope at 1.4 T. (b) and (c) Magnetic field dependence of critical current density $J_c$ and pinning force density $F_p$ in double-logarithmic representation, corresponding to Figure 5a,b. The limits of power law regions in these two figures are between 1 and 1.5 T. The intersection of lines in (c) can serve as a measure for the pinning force per flux line length of the TJs.

## 3. The quasi-hexagonal grain in the MgO template

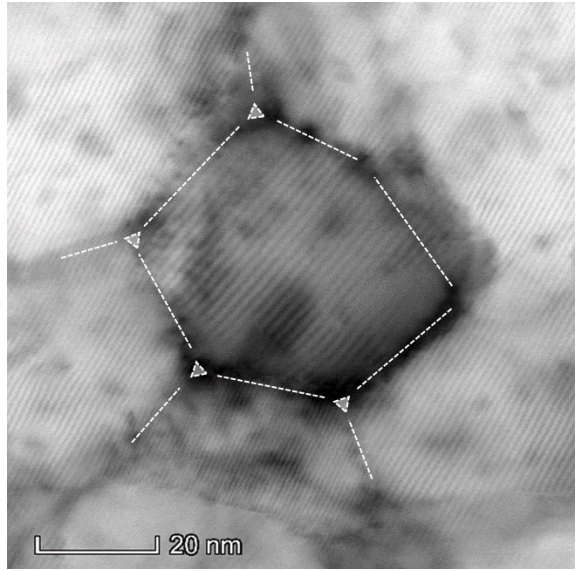

Figure S4. Plan-view high-resolution bright-filed STEM image of the IBAD-MgO template. A quasi-hexagonal grain is clearly visible.

## 4. Crystallographic orientation and Euler angles

Figure S5 (a) shows the schematic diagram of the relationship between the crystal coordinate system $a$, $b$, $c$ ([100], [010], [001]) and the original fixed coordinate system $X$, $Y$, $Z$, descripted by Euler angles, $α$, $β$, $γ$.[11] Euler angles are one possible set of mathematical parameters used to describe crystal orientation. At the beginning, the green line of nodes needs to be defined as the intersection of the $ab$-plane and $XY$-plane, displayed as $N$-axis. Then $α$, $β$, $γ$ are defined as the angles between the $X$-axis and $N$-axis, $Z$-axis and $c$-axis, $N$-axis and $a$-axis, respectively. In terms of definition by crystal intrinsic rotations, these three angles also characterize the three elemental rotations of the crystal intrinsic system starting at the $XYZ$ system: I-rotation around $Z$-axis by angle $α$, II-rotation around $N$-axis by angle $β$, and III-rotation around $c$-axis by angle $γ$, respectively.

Based on the aforementioned, in Figure S5 (b), the angles $γ$ and $β$ can be regarded as crystal rotation around $c$-axis (in-plane rotation angle, $γ$) and around an axis within the $ab$-plane (out-

S-5

of-plane rotation angle, β), respectively. In addition, the symmetry of crystalline materials has to be considered. There are 24 different sets of Euler angles (α, β, γ) for one crystal orientation in a cubic system (8 sets for crystals with tetragonal symmetry). Only after unifying the orientation data of each point of the sample into the same set of Euler angles, can these orientation data possess comparability with each other in individual Euler angle α, β or γ. Therefore, all orientation Euler angles in this report were unified into the same set before the in-plane and out-of-plane crystal rotation angles (γ and β) were investigated based on SPED orientation maps. In addition, Figure S5 (b) shows that the values of β are around 90° and γ around 0°. So, for the sake of comparison, 90° was subtracted from all values of β for Figure 7 of this report.

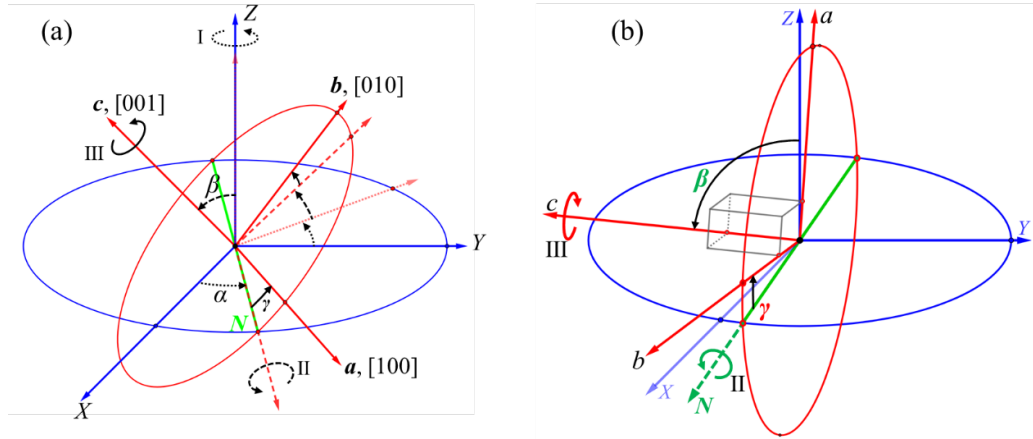

Figure S5. (a) Schematic diagram of the orientation relationship between crystal intrinsic system *a*, *b*, *c* ([100], [010], [001], red) and reference coordinate system *X*, *Y*, *Z* (blue), described by Euler angles α, β, γ. (b) The crystal orientation angles, β and γ of the tetragonal crystal lattice of this cross-section TEM foil (*a, b, c*) in the microscope coordinate system (*X, Y, Z*). In this state, γ is regarded as in-plane crystal rotation angle, and β as out-of-plane crystal rotation angle.

## 5. Detailed investigations of crystal rotation of IBAD-MgO substrate

Figure S6 (b) illustrates nicely that the statistical out-of-plane misorientation $\Delta\psi_\beta^{MgO}$ is decreasing through the thickness of MgO as approaching the NdFeAs(O,F)/MgO interface.

While the statistical in-plane misorientation $\Delta\psi_\gamma^{MgO}$, Figure S6 (c), decreased at first but increased back when close to the interface, which might be ascribed to the crystal lattice in-plane distortion caused by lattice mismatch between NdFeAs(O,F) and MgO. Additionally, by comparing $\Delta\psi^{SC}$ ($\Delta\psi_\beta^{SC}$ = 1.7° and $\Delta\psi_\gamma^{SC}$ = 2.4°) with $\Delta\psi^{MgO}$ at -15 nm depth ($\Delta\psi_\beta^{MgO}$(-15) = 1.0° and $\Delta\psi_\gamma^{MgO}$(-15) = 2.2°), we realize the degree of texture of epitaxial NdFeAs(O,F) film is slightly lower than of the MgO substrate for both out-of-plane and in-plane. However, in Figure S6 (a) and (d), the profile of NdFeAs(O,F) is still similar to that of MgO at -15 nm depth, which further evidences the orientation dominance of the MgO substrate on the NdFeAs(O,F) film. Consequently, it shows at the nanoscale that the optimization of the grain boundaries of epitaxial superconducting films can be achieved to some extent by engineering the thickness of the MgO layer.

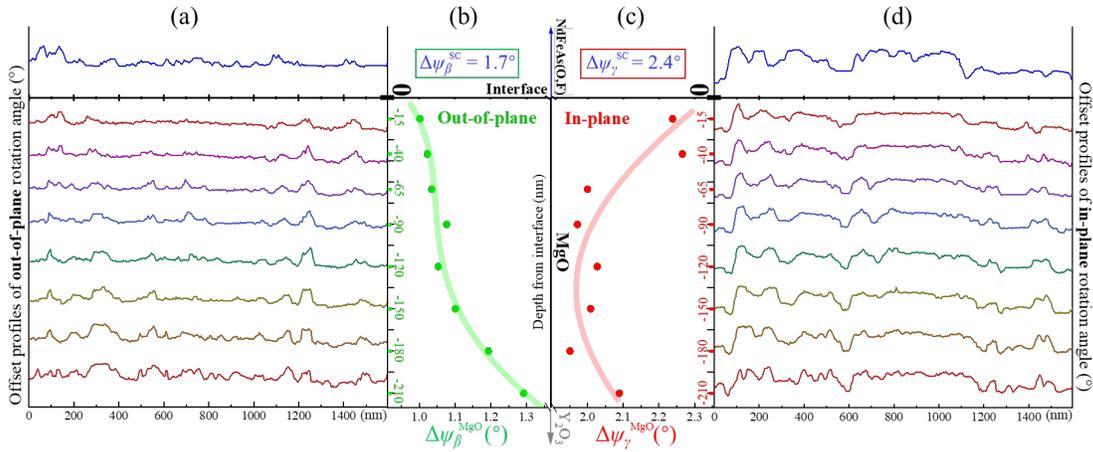

Figure S6. (a) and (d) The offset angular profiles (from the same set of data as Figure 7) of out-of-plane and in-plane crystal rotation of MgO substrate at different depths from the NdFeAs(O,F)/MgO interface. The angular profiles of NdFeAs(O,F) superconducting (SC) film are also displayed on top. (b) and (c) The full width at half maximum (FWHM, $\Delta\psi$) values of the distribution histograms (similar to Figure 7 (d)) corresponding to the offset profiles of MgO in (a) and (d). They were counted from all 10 sets of orientation data of which one in each depth is shown, and regarded as the statistical out-of-plane (green) and in-plane (red) misorientation angle ($\Delta\psi_\beta^{MgO}$ and $\Delta\psi_\gamma^{MgO}$). The data points of $\Delta\psi^{MgO}$ depending on the depth are fitted by a 4$^{th}$ order polynomial function. The out-of-plane and in-plane $\Delta\psi$ of NdFeAs(O,F) film, $\Delta\psi_\beta^{SC}$ and $\Delta\psi_\gamma^{SC}$ are also shown for comparison.

## 6. Out-of-plane misorientation angles of this NdFeAs(O,F) film

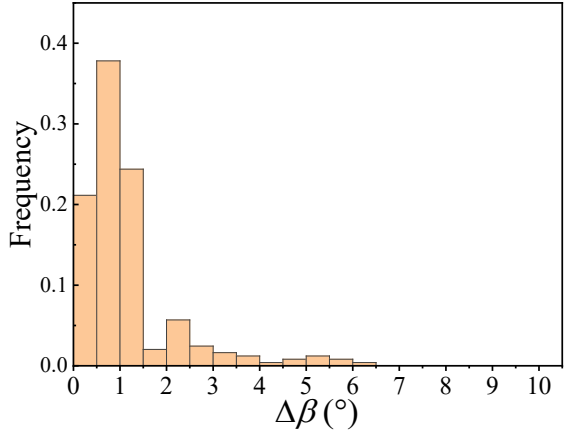

Figure S7. The histogram of out-of-plane misorientation angles between adjacent grains, $\Delta\beta$. More than 90% of $\Delta\beta$ in this epitaxial NdFeAs(O,F) film were less than 2.5°. The maximum $\Delta\beta$ was 6.3°.

## 7. Main symbols in this report

$\Delta g$ — Misorientation angle between adjacent grains, /deg. (°).

$\gamma$ / $\beta$ — In-plane / out-of-plane crystal rotation angle of every data point, /deg. (°). This magnitude is the absolute value in the certain microscope coordinate system, so that takes both positive and negative.

$\Delta\psi$ ($\Delta\psi_{\gamma}^{SC}$, $\Delta\psi_{\beta}^{SC}$, $\Delta\psi_{\gamma}^{MgO}$, $\Delta\psi_{\beta}^{MgO}$) — Full width at half maximum of the histogram of in-plane ($\gamma$) and out-of-plane ($\beta$) rotation angle distribution counting from all 10 sets of data, /deg. (°), for the NdFeAs(O,F) superconducting (SC) film and the MgO substrate.

$\Delta\gamma$ / $\Delta\beta$ — In-plane / out-of-plane misorientation angle between adjacent grains, /deg. (°). The $\Delta\gamma$ is defined for comparison with the critical angle, $\theta_c$, regarding the [001]-tilt.

$d_g$ — The distance between adjacent grain boundaries measured from the cross-section TEM foils, /nm.

*d* — Mean grain size (diameter), /nm. This variable is defined for calculation.

## References

[1] Rauch, E. F.; Portillo, J.; Nicolopoulos, S.; Bultreys, D.; Rouvimov, S.; Moeck, P. Automated Nanocrystal Orientation and Phase Mapping in the Transmission Electron Microscope on the Basis of Precession Electron Diffraction. *Zeitschrift fur Krist.* **2010**, *225*, 103–109.

[2] Wu, G.; Zaefferer, S. Advances in TEM Orientation Microscopy by Combination of Dark-Field Conical Scanning and Improved Image Matching. *Ultramicroscopy* **2009**, *109*, 1317–1325.

[3] Barnard, J. S.; Johnstone, D. N.; Midgley, P. A. High-Resolution Scanning Precession Electron Diffraction: Alignment and Spatial Resolution. *Ultramicroscopy* **2017**, *174*, 79–88.

[4] Vincent, R.; Midgley, P. A. Double Conical Beam-Rocking System for Measurement of Integrated Electron Diffraction Intensities. *Ultramicroscopy* **1994**, *53*, 271–282.

[5] Sinkler, W.; Marks, L. D. Characteristics of Precession Electron Diffraction Intensities from Dynamical Simulations. *Zeitschrift fur Krist.* **2010**, *225* (2–3), 47–55.

[6] White, T. A.; Eggeman, A. S.; Midgley, P. A. Is Precession Electron Diffraction Kinematical? Part I: "Phase-Scrambling" Multislice Simulations. *Ultramicroscopy* **2010**, *110*, 763–770.

[7] Eggeman, A. S.; White, T. A.; Midgley, P. A. Is Precession Electron Diffraction Kinematical? Part II. A Practical Method to Determine the Optimum Precession Angle. *Ultramicroscopy* **2010**, *110*, 771–777.

[8] Midgley, P. A.; Sleight, M. E.; Vincent, R. The Structure of a Metastable Au-Sn Phase Determined by Convergent Beam Electron Diffraction. *J. Solid State Chem.* **1996**, *124*, 132–142.

[9] Sinkler, W.; Marks, L. D.; Edwards, D. D.; Mason, T. O.; Poeppelmeier, K. R.; Hu, Z.; Jorgensen, J. D. Determination of Oxygen Atomic Positions in a Ga-In-Sn-O Ceramic Using Direct Methods and Electron Diffraction. *J. Solid State Chem.* **1998**, *136*, 145–149.

[10] Gjønnes, J.; Hansen, V.; Berg, B. S.; Runde, P.; Cheng, Y. F.; Gjønnes, K.; Dorset, D. L.; Gilmore, C. J. Structure Model for the Phase AlmFe Derived from Three-Dimensional Electron Diffraction Intensity Data Collected by a Precession Technique. Comparison with Convergent-Beam Diffraction. *Acta Crystallogr. Sect. A Found. Crystallogr.* **1998**, *54*, 306–319.

[11] Engler, O.; Randle, V. *Introduction to Texture Analysis*; CRC press, 2009.



\end{document}